\documentclass[preprint2]{aastex}

\usepackage{rotating}
\slugcomment{}
\shorttitle{Red Nuggets at $z\sim 1.5$}
\shortauthors{Damjanov et al.}
\begin{document}
\title{Red Nuggets at $z\sim 1.5$: Compact passive galaxies and the formation of the Kormendy Relation}

\author{\medskip 
Ivana Damjanov\altaffilmark{\dag}, 
Patrick J. McCarthy\altaffilmark{\ddag}, 
Roberto G. Abraham\altaffilmark{\dag}, 
Karl Glazebrook\altaffilmark{\P}, 
Haojing Yan\altaffilmark{\ddag}, 
Erin Mentuch\altaffilmark{\dag},
Damien Le Borgne\altaffilmark{\diamond}, 
Sandra Savaglio\altaffilmark{\S}, 
David Crampton\altaffilmark{\circ}, 
Richard Murowinski\altaffilmark{\circ}, 
St{\'e}phanie Juneau\altaffilmark{\sun}, 
R. G. Carlberg\altaffilmark{\dag},
Inger J{\o}rgensen\altaffilmark{\triangleright}, 
Kathy Roth\altaffilmark{\triangleright}, 
Hsiao-Wen Chen\altaffilmark{\wedge}, and 
Ronald O. Marzke\altaffilmark{\star}}

\altaffiltext{\dag}{Department of Astronomy \& Astrophysics, University of Toronto, 50 St. George Street, Toronto, ON, M5S~3H4}

\altaffiltext{\ddag}{Observatories of the Carnegie Institution of Washington,813 Santa Barbara Street, Pasadena, CA 91101}

\altaffiltext{\P}{Centre for Astrophysics and Supercomputing, Swinburne University of Technology, 1 Alfred St, Hawthorn, Victoria 3122, Australia}

\altaffiltext{\S}{Max-Planck-Institut f\"ur extraterrestrische Physik, Garching, Germany}

\altaffiltext{$\circ$}{Herzberg Institute of Astrophysics, National Research Council, 5071 West Saanich Road, Victoria, British Columbia, V9E~2E7, Canada.} 

\altaffiltext{$\sun$}{Department of Astronomy \slash  Steward Observatory, University of Arizona, 933 N Cherry Ave., Rm. N204, Tucson AZ 85721-0065}

\altaffiltext{$\diamond$}{DSM\slash ~DAPNIA\slash ~Service d'Astrophysique, CEA\slash ~SACLAY, 91191 Gif-sur-Yvette Cedex, France}

\altaffiltext{$\triangleright$}{Gemini Observatory, Hilo, HI 96720}

\altaffiltext{$\wedge$}{The Department of Astonomy and Astrophysics, University of Chicago, 5640 S. Ellis Ave, Chicago, IL 60637}

\altaffiltext{$\star$}{Dept. of Physics and Astronomy, San Francisco State University, 1600 Holloway Avenue, San Francisco, CA 94132}

\defcitealias{abr04}{Paper~I}
\defcitealias{gla04}{Paper~III}
\defcitealias{mcc04}{Paper~IV}
\defcitealias{abr07}{Paper~VIII}

\begin{abstract}
We present the results of NICMOS imaging of a sample of 19  high mass passively evolving galaxies 
with $1.2<z<2$, taken primarily from the Gemini Deep Deep Survey (GDDS).
Around $80\%$ of galaxies in our GDDS sample have spectra dominated by stars with 
ages $\gtrsim1$~Gyr.  Our rest-frame $R$-band images show
that most of these objects have compact regular morphologies which follow
the classical R$^{1\slash 4}$~law.  These galaxies scatter along a tight sequence in the size vs. 
surface brightness parameter space which defines the Kormendy relation. Around one-third ($3\slash 10$)
of the massive red objects in the GDDS sample are extraordinarily compact, with effective radii under one kiloparsec. 
Our NICMOS observations allow the detection of such systems more robustly
than is possible with optical (rest-frame UV) data, and while similar systems
have been seen at $z\ga2$, this
is the first time such systems have been detected in a rest-frame optical survey
at $1.2<z<2$.
We refer to these compact galaxies as `red nuggets', and note that similarly compact massive galaxies
are completely absent in the nearby Universe.
We introduce a new `stellar mass Kormendy relation' 
(stellar mass density vs size) which we use to single out the effects of size 
evolution from those of luminosity and color evolution in stellar populations.
The $1< z < 2$ passive galaxies have mass densities that are an order
of magnitude larger then early type galaxies today and are comparable to the compact distant red
galaxies at $2 < z < 3$. 
We briefly consider mechanisms for size evolution in contemporary models
focusing on equal-mass mergers and adiabatic expansion driven by stellar mass loss. 
Neither of these mechanisms appears able to transform the high-redshift Kormendy relation
into its local counterpart, leaving the origin and fate of these compact `red nuggets' unresolved.

\bigskip
\end{abstract}


\section{Introduction}

The formation mechanism of elliptical galaxies has long been controversial and 
remains a key test of more general galaxy formation models. The original `nature' (\citet{elbs62} monolithic collapse) vs. `nurture' (formation through mergers, \citep[e.g.,][]{schweizer87,searle78,tt72}
debate is still with us, but is now set in a  $\Lambda$CDM cosmological 
context which attempts to connect the stellar component of galaxies to an underlying
evolutionary picture for the clustering of dark matter halos. Testing this model requires
studying the evolution of galaxies over a large redshift range.

A wide range of selection 
techniques have been effective in selecting galaxies in various redshift ranges 
on the basis of their current star formation rates (e.g. Lyman break galaxies, sub-mm sources
 etc), or from the spectral signatures of passively evolving old stellar populations 
(e.g., extremely red objects (EROs) and other color selections). 
The most massive local elliptical galaxies have the oldest stellar populations 
\citep{ght84}, so identifying the progenitors of local early-type
galaxies within the high-redshift galaxy population is of particular interest.
There is a consensus that the mass density in the 
red sequence is evolving strongly in the $1<z<2$ range (\citealp[GDDS Paper VIII,][]{abr07}; \citealp[GDDS Paper III,][]{gla04}; \citealp{fon04,ru03}), 
a process that continues at redshifts below unity as well \citep{fab07,bell04}, although the
magnitude of the evolution is uncertain \citep{brown07,chen03}. 
Massive morphologically-confirmed
elliptical galaxies have been found up to 
$z=2$ (\citealp[GDDS Paper IV,][] {mcc04}; \citealp{cim04}) with
spectra consistent with formation epochs up to 
$z>5$. These observations were in  
in direct contradiction with early  $\Lambda$CDM
models where stellar mass assembly traced the build
up of cold dark matter haloes, 
although additional feedback mechanisms
on the baryons have more recently been able to better account 
for this \citep[e.g.,][]{KJS06}. A
complication recently added to this picture is the observation that
the space density of 
ellipticals is found to evolve strongly over $1<z<2$ \citepalias{abr07} even while their 
stellar populations evolve weakly, suggesting that one must
be careful to decouple morphological evolution from evolution of the
underlying stellar populations. This is seen at higher redshifts
also, where
the paucity of passively evolving galaxies at $z>2$ in deep $J-K$ and 
3.5~$\mu$m selected samples \citep{kr06,la05,cim02} shows that the
$assembly$ epoch for the red sequence may be decoupled from the epoch
of the earliest star formation.  Studies of star formation history and morphology can only go so
far in unraveling the puzzle of galaxy formation; dynamical and chemical probes are 
needed to connect progenitors to descendants.  Clustering signatures offer one dynamical
approach to connecting progenitors to descendants and the strong clustering of the 
passive red galaxies \citep{dad05a,dad04,brown03,mcc01} strongly suggest 
that they are linked to today's massive ellipitical galaxies. 

Theoretical attempts to explain these observations
have resulted in greatly improved  $\Lambda$CDM models which 
decouple mass assembly from this Ôstellar population downsizingÕ. An example is the semi-analytic model of \citet{dlb07}. Here the small ellipticals and their stars form early 
by disc mergers. Massive ellipticals can then grow bigger and more numerous at late times 
through dissipationless or ÔdryÕ merging. This may even have been observed \citep{bell06a} though there is some disagreement as to whether the $\Lambda$CDM  merger rate is high 
enough \citep{bte07}. 
At this stage it is perhaps fair to say that dry merging simulations show that it does not
disrupt elliptical scaling relations  \citep{bk06,bk05,gva03} as one might naively expect. However
only a limited number of simulations of this process have been done and they
have not yet been incorporated into cosmological models in a detailed way
such that they can be compared with data (e.g., numbers, sizes and masses of galaxies).
Further it is not clear that a dry
merging hierarchy consistent with cosmological downsizing can also be made consistent
with the evolving
mass-metallicity relation \citep{pip08}. 
A contrasting picture is painted by \citet{na07} using a 
SPH model of individual systems. They argue for a formation mode dominated by something 
very close to early monolithic collapse, but in a  $\Lambda$CDM  cosmological context, with mergers 
(along with accretion) playing only a minor role in stellar mass growth at late times. 

High spatial resolution studies of the morphologies and structures of passive galaxies offer one approach to 
gauging the importance of recent major merger events. 
A number of studies with the {\em Hubble 
Space Telescope} (HST) have shown that half or more of red galaxies in color-selected 
samples have simple early type morphologies. Most of these studies are confined to 
redshifts of $\sim1.5$ and less, and the early-type fraction varies from $\sim50\%$ 
to $70\%$ \citep{mou04,yan03}. At higher redshifts a significant fraction
of the red galaxies appear to be discs (e.g., \citetalias{abr07}, \citealp{fon04}). Understanding the connection
between these two classes of objects naturally focuses on the
importance of mergers, since nearly equal-mass mergers are thought to transform discs into spheroids.
Mergers, both gas-rich and dissipationless, 
are also thought to be important in the growth of the red sequence  and 
evidence, both direct and indirect, supports that this is occuring at intermediate and low redshifts \citep[e.g.,][and the references therein]{bell06b}.
It appears that much of the high-redshift merging activity may
be of the dissipationless variety where the main effect of merging is to reorganize existing
stellar population {\em without} triggering new star formation. It is
difficult to envision how this might operate unless the merging systems
are themselves gas-poor, which is not generally expected \citep{van05}. In any case,
the signatures of such `dry' mergers are difficult to detect at high redshifts.

Recently, several imaging studies have shown that red galaxies at $z>1$ appear smaller than their likely present-day descendants
with the same stellar mass \citep{lon07,cim08}. The implications of these observations are seen most clearly in the 
structural and dynamical scaling relations, 
the Fundamental Plane and its projections (the Faber-Jackson (1976) and Kormendy (1977) 
relations). In the present paper we explore the nature of the Kormendy relation,
(mean surface brightness within the effective radius, $\langle\mu\rangle_e$,
versus effective radius, R$_e$). This is
the most observationally accessible projection of the fundamental plane at high-redshift.
Our analysis spans the redshift range
$1.2 < z < 2$ using HST NICMOS observations of a sample of quiescent high-redshift galaxies
taken mainly from the  Gemini Deep Deep Survey \citep[GDDS Paper I,][]{abr04}.
We present NICMOS F160W images for ten of the twenty $z>1.3$ passive red galaxies 
from \citetalias{mcc04}. These systems all have spectra dominated by old stellar populations. 
This extends to higher redshifts ($z>1.7$) than the earlier NICMOS work of \citet{lon07} from the Munich Near-IR Cluster Survey \citep[MUNICS,][]{dr01}. We 
also independently analyze the archival NICMOS data of \citet{lon07} in 
the redshift range $1.2 < z < 1.7$ to supplement our sample and confirm their
findings. At the higher redshifts previous findings of compact galaxies were based on optical data obtained with the Advanced Camera for Surveys (ACS) onboard HST \citep{cim08}. 
Our use of NICMOS allows us to more robustly show that the old components in the galaxies are truly compact. 
Finally, we are able to unify the optical and infrared work by introducing a new `stellar mass Kormendy relation' 
which we use to  better quantify evolution in the sizes of early-type galaxies as a function of stellar mass over the redshift range $1<z<2$. 
We briefly examine the likelihood that dry mergers explain such size evolution, and examine whether an
alternative process, adiabatic expansion, might be more important.
We describe the observations in section~\ref{obs}, our analysis in section~\ref{analysis}, 
and present our results in section~\ref{res}. In section~\ref{disc} we discuss the
implications of our observations for simple models for galaxy size
growth. Throughout we use standard cosmological parameters; H$_0=70$~km~s$^{-1}$~Mpc$^{-1}$, 
$\Omega_\textrm{m}=0.3$, $\Omega_\Lambda=0.7$. Unless stated
otherwise, all magnitudes are based on the AB system.

\section{Description of the Observations}\label{obs}
\subsection{Sample definition}\label{sdef} 

Our sample of galaxies was taken mainly from the GDDS, crafted to sample 
the galaxy population in the critical $1<z<2$ interval with an emphasis on red 
galaxies \citepalias{abr04}. While modest in area (120 square arcminutes), the survey is spread over 
four independent and representative sightlines. Redshifts for $\sim 300$ galaxies brighter 
than $I(\textrm{Vega}) = 24.5$ were obtained from 30-hour long integrations using 
the GMOS spectrometer on Gemini North. This magnitude limit corresponds to the stellar mass of 
$2.5\times 10^{10}$~M$_\sun$ for a galaxy with the redshift of formation $z_f=10$ and maximally old stellar population observed at redshift $z=1.5$ \citepalias{gla04} . 
We classified the galaxies on the basis of their spectra, depending on whether they were dominated by active star formation, stars 
older than  $\sim1$~Gyr, intermediate age ($0.3-1$~Gyr) populations, or a mix of these 
types. Of the 302 galaxies with redshifts, 47 have spectra dominated by 
old stars, and twenty of these lie at redshifts beyond 1.3. Spectra of these twenty
galaxies and estimates of their ages and formation redshifts are presented in 
\citetalias{mcc04}. Deep $I$-band images of the GDDS galaxies at $z<1.7$ with the 
ACS on HST reveal that the correlation between spectral 
type, and hence stellar content, and morphological class seen at present is strong at these 
redshifts. Nearly all of the GDDS galaxies with passive spectral classes have compact 
morphologies consistent with early Hubble types, while the actively star forming 
galaxies have a morphologies that range from simple disks to complex structures indicative 
of ongoing mergers. The GDDS galaxies discussed in this paper are a subset of the GDDS 
galaxies having spectra dominated by old stars (class ``001'' from \citetalias{abr04}) 
and $z>1.3$.  The key properties of this sample are given in Table~\ref{tab1}.

Our primary sample of ten galaxies is drawn from the GDDS and determined by the number of
available orbits and the desired depth of NICMOS imaging. The targets were selected randomly,
with the exception of the two (12-5869 and 12-5592) that could be covered in a single
pointing. We also analyzed archival data from
the MUNICS survey for nine additional galaxies with properties similar to those of
our GDDS sample. \citet{lon05} analyzed spectrophotometric data set for these galaxies from the near-infared spectroscopic  follow-up of a complete sample of bright ($K<18.5$) EROs ($R-K>5.3$) selected from the
MUNICS survey\footnote{This is actually a blank field survey, the intention was to {\em find} high-$z$ clusters from deep wide-field near-IR imaging.}. Low resolution spectroscopic and photometric data  revealed
stellar masses greater than $10^{11}$~M$_{\sun}$ and dominant old stellar population for all objects in the sample (see Table~\ref{tab2}). 
As will be described below, this additional data provided us with a useful check of our methodology by allowing us to compare results from our analysis pipeline against those published 
in \citet{lon07}.

\begin{deluxetable}{lccc}
\tablecolumns{4}
\tablewidth{0pc}
\tablecaption{Properties of the ten galaxies in GDDS sample\label{tab1}}
\tabletypesize{}
\tablehead{
\colhead{ID} & \colhead{$z$} & \colhead{Mass\tablenotemark{a}} & \colhead{Age\tablenotemark{b}} \\
\colhead{} & \colhead{} & \colhead{[$10^{11}$~M$_{\sun}$]} &  \colhead{[Gyr]}} 
\startdata
12-5592\dotfill&1.623&1.16 $\pm$ 0.27&$1.1^{+0.3}_{-0.4}$\\
12-5869\dotfill&1.51 &3.14 $\pm$ 0.43 &1.2$^{+0.6}_{-0.2}$\\ 
12-6072\dotfill&1.576&0.59 $\pm$ 0.27&$1.6^{+2.1}_{-1.3}$\\
12-8025\dotfill&1.397&1.25 $\pm$ 0.39&$0.8^{+0.6}_{-0.1}$\\
12-8895\dotfill&1.646&3.18 $\pm$ 0.44&$2.5^{+0.3}_{-0.3}$\\
15-4367\dotfill&1.725&0.56 $\pm$ 0.15&$2.1^{+0.4}_{-0.9}$\\
15-5005\dotfill&1.845&0.67 $\pm$ 0.24&$0.5^{+0.7}_{-0.1}$\\
15-7543\dotfill&1.801&1.06 $\pm$ 0.30&$0.9^{+0.5}_{-0.2}$\\
22-0189\dotfill&1.49& 2.85 $\pm$ 0.98&$3.0^{+0.7}_{-0.2}$\\
22-1983\dotfill&1.488&1.34 $\pm$ 0.53&$1.1^{+3.1}_{-0.5}$\\
\enddata
\tablenotetext{a}{\ GDDS mass estimates are based on the \citet{bal03} IMF,  and taken from \citetalias{gla04}}
\tablenotetext{b}{\ Minimum galaxy ages from \citetalias{mcc04}}
\end{deluxetable}

\begin{deluxetable}{lccc}
\tablecolumns{4}
\tablewidth{0pc}
\tablecaption{Properties of six massive galaxies in MUNICS sample\tablenotemark{a}\label{tab2}}
\tabletypesize{}
\tablehead{
\colhead{ID} & \colhead{$z$} & \colhead{Mass\tablenotemark{b}} & \colhead{Age} \\
\colhead{} & \colhead{} & \colhead{[$10^{11}$~M$_{\sun}$]} &  \colhead{[Gyr]}} 
\startdata
S2F5\_109&1.22&5.94 $\pm$ 0.95 &1.7 $\pm$ 0.3\\
S7F5\_254&1.22&4.68 $\pm$ 0.16&5.0 $\pm$ 0.1\\
S2F1\_357&1.34&4.65 $\pm$ 0.40&4.0 $\pm$ 0.1\\
S2F1\_389&1.40&2.15 $\pm$ 0.86&3.0 $\pm$ 0.5\\
S2F1\_511&1.40&2.07 $\pm$ 0.89&1.3 $\pm$ 0.3\\
S2F1\_142&1.43&4.06 $\pm$ 0.94&2.2 $\pm$ 0.2\\
S7F5\_045&1.45&3.58 $\pm$ 1.10&1.7 $\pm$ 0.3\\
S2F1\_633&1.45&3.52 $\pm$ 0.51&4.0 $\pm$ 0.5\\
S2F1\_443&1.70&3.58 $\pm$ 1.48&3.5 $\pm$ 0.3\\
\enddata
\tablenotetext{a}{\ from \citet{lon07}} 
\tablenotetext{b}{\ MUNICS mass estimates are taken from \citet[][Salpeter IMF]{lon05}, and transformed to \citet{bal03} IMF following the relation given in \citetalias{gla04}}
\end{deluxetable}

\subsection{NICMOS Observations}

The ten GDDS galaxies were observed with Camera 3 on NICMOS using 
the F160W filter. Each individual exposure was 896 seconds in duration with multiple samples using 
the STEP64 read pattern. A single orbit contained three exposures and we observed 
each target over four HST orbits for a total integration time of 10740 seconds. Two
of the fields overlapped and the images for targets 12-5869 and 12-5592 have twice the
exposure time of the others. These objects are discussed in detail in 
\citet{mcc07}.  We dithered in non-integer pixel steps between each 
exposure. The individual frames were dark corrected, sky subtracted and combined using 
the DRIZZLE algorithm \citep{fh02} with a final pixel size of $0\farcs12$. 
Residual sky levels in the final mosaics were derived from Gaussian fits to a histogram 
of sky values and were subtracted.

As noted above, we also re-analyised nine galaxies from the MUNICS
sample of red galaxies described in \citet{lon07}. The MUNICS data set was obtained using Camera 
2 on NICMOS, and is thus more finely sampled, and somewhat shallower, than our NIC3 images.
As described below, analyzing this NIC2 data allowed us to explore, and ultimately rule out, 
the possibility that the coarser sampling of our NIC3 data might lead to poor model fits and
spurious sizes.  We retrieved the pipeline-processed individual NIC2 images from the
HST archive. We then corrected each image for residual pedestal effects and combined them
into mosaics using the DRIZZLE algorithm with a final pixel size of $0\farcs05$.
The properties of the nine galaxies in this sample are summarized in Table~2.

\section{Analysis}\label{analysis}

\subsection{Surface brightness profiles}

\begin{figure*}[htp]
\epsscale{1}\plotone{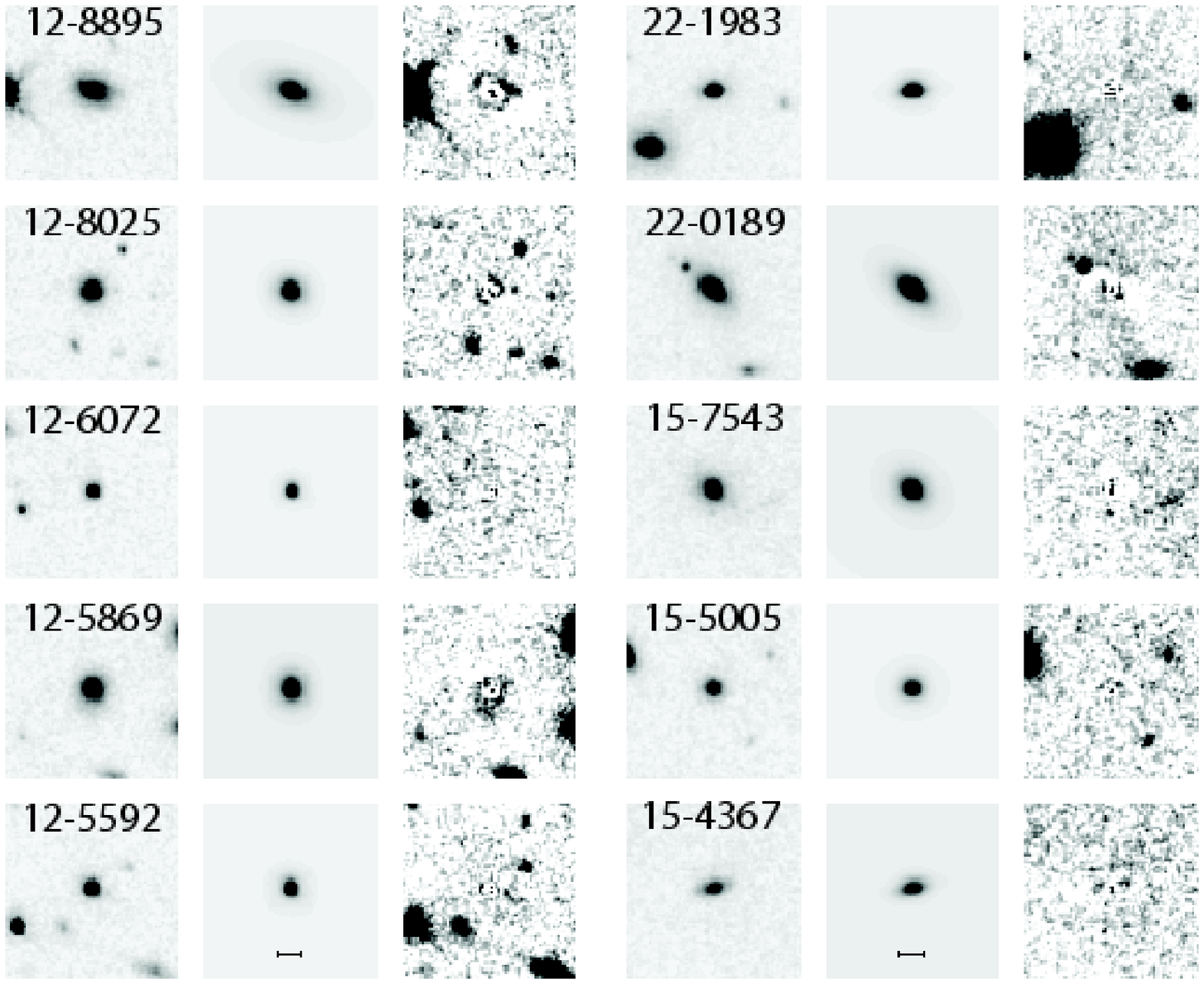}
\caption{NIC3 images and the results of our 2D fitting with \texttt{Galfit} for 
our sample of 10 GDDS galaxies with $1.3 < z < 2$ and spectra dominated by old stars. The three columns present 
the drizzled F160W image, the best fitting R$^{1\slash 4}$ model, and the residuals. The residual images
have been scaled by a factor of 10 compared to the data and models to bring out faint
features.  The bars at the bottom are one arcsecond in length. }
\label{f1}
\end{figure*}

\begin{figure*}[htp]
\epsscale{1}\plotone{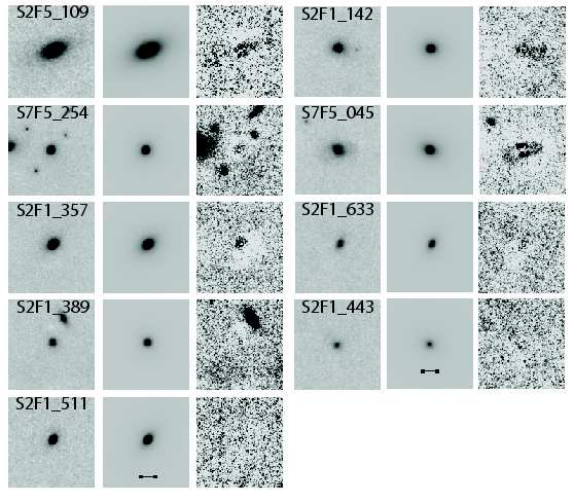}
\caption{NIC2 images and the results of our 2D fitting with \texttt{Galfit}  
of the six galaxies from \citet{lon07}.
The three columns present the galaxy, the best fitting R$^{1\slash 4}$ model, and the residuals. The residual images 
have been scaled by a factor of 10 compared to the data and models to bring out
faint features. The bars at the bottom are one arcsecond in length. }
\label{f2}
\end{figure*}

Using the \texttt{Galfit} software package \citep{pe02}, we derived 
two-dimensional (2D) surface brightness profiles by fitting synthetic galaxy images to
our data using a 
range of surface brightness profiles, ellipticities and orientations. A series of models 
were constructed using exponential surface brightness profiles, de Vaucouleurs R$^{1\slash 4}$
profiles and the more general R$^{1\slash  n}$ S\'ersic profiles.
We did not consider more general fitting laws due to the relatively small range of radii ($0\farcs12-2\arcsec$, or
$1-17$~kpc at $z=1.5$)  covered by our observations. Models with a range of scale lengths and eccentricities 
were convolved with the Point Spread Function (PSF) of the observations and subtracted 
from the NICMOS images.  We used PSFs derived from well-detected 
unsaturated stars in each NIC3 field rather than the TinyTim simulations as we found the 
former produced better fits.  The residuals were computed and the model parameters were 
iterated to minimize the square of the residuals within the box 
of $8\farcs4\times8\farcs4$ centered on each galaxy. The initial guess for the
centroid was the position of the highest intensity pixel within the fitting box, and the
total magnitude was estimated according to the total intensity confined in this box.
Both initial guesses were made after masking out of the neighbouring sources. 
The root mean square (RMS) image
was used to give relative weights to the background pixels during the fitting. By using
different stars the width of the NIC3 PSF was
allowed to vary to include the effects of spatial and temporal variations in the NIC3
PSF. Changing the PSF had very little impact on the derived effective radii in
all cases. The best-fit models for all galaxies in the sample are presented in
Figure~\ref{f1} (middle column) along with the residual images
(last column). Parameters of the best-fit R$^{1\slash 4}$ and R$^{1\slash n}$
profiles for each galaxy are given in Table~\ref{tab3}. 
 The listed minima of reduced $\chi^2$ are well below unity, 
suggesting that the flux uncertainties introduced by the RMS images are overestimated.  
We performed the same morphological analysis on the MUNICS galaxies \citep{lon07}. 
The NIC2 PSF used for modeling 2D profiles of these objects was derived from the TinyTim simulations. 
The resulting best-fit R$^{1\slash 4}$ profiles are graphically illustrated 
in Figure~\ref{f2}. The parameters obtained are listed in Table~\ref{tab4}, along with the results from \citet{lon07} for comparison. 
The reduced $\chi^2$ are again below unity, but the values obtained for our best fit are very similar to the ones obtained for \citet{lon07} parameters, except for the total F160W magnitudes where the difference is greater then $1\sigma$. The reasons for this discrepancy may be the simulated PSF we used for 2D fitting and the different methods applied for background subtraction.  Also, resulting R$^{1\slash 4}$ fit  effective radius R$_e$ and surface brightness $\langle\mu\rangle_e$  for objects S2F1\_142, S7F5\_45, S2F1\_633, and S2F1\_443 differ for more then $1\sigma$ from the previously reported ones. When fitted with R$^{1\slash n}$ profiles, the best fits for the three of these objects -  S2F1\_142,  S2F1\_633, and S7F5\_45 - have lower indices $n$ than listed in \citet{lon07} - 2 instead of 3.5, 2.5 instead of 4.1, and 1.5 instead of 2, respectively. On the other hand, the best fit R$^{1\slash n}$ profile for S2F1\_443 has index $n=2.8$, higher than $n=1.9$ reported by \citet{lon07}. For the rest of the MUNICS sample the difference in the goodness of fit for R$^{1\slash 4}$ profile between our and \citet{lon07} analysis is $\Delta(\chi^2)\lesssim0.2$.

\begin{figure*}[htp]
\epsscale{2}\plotone{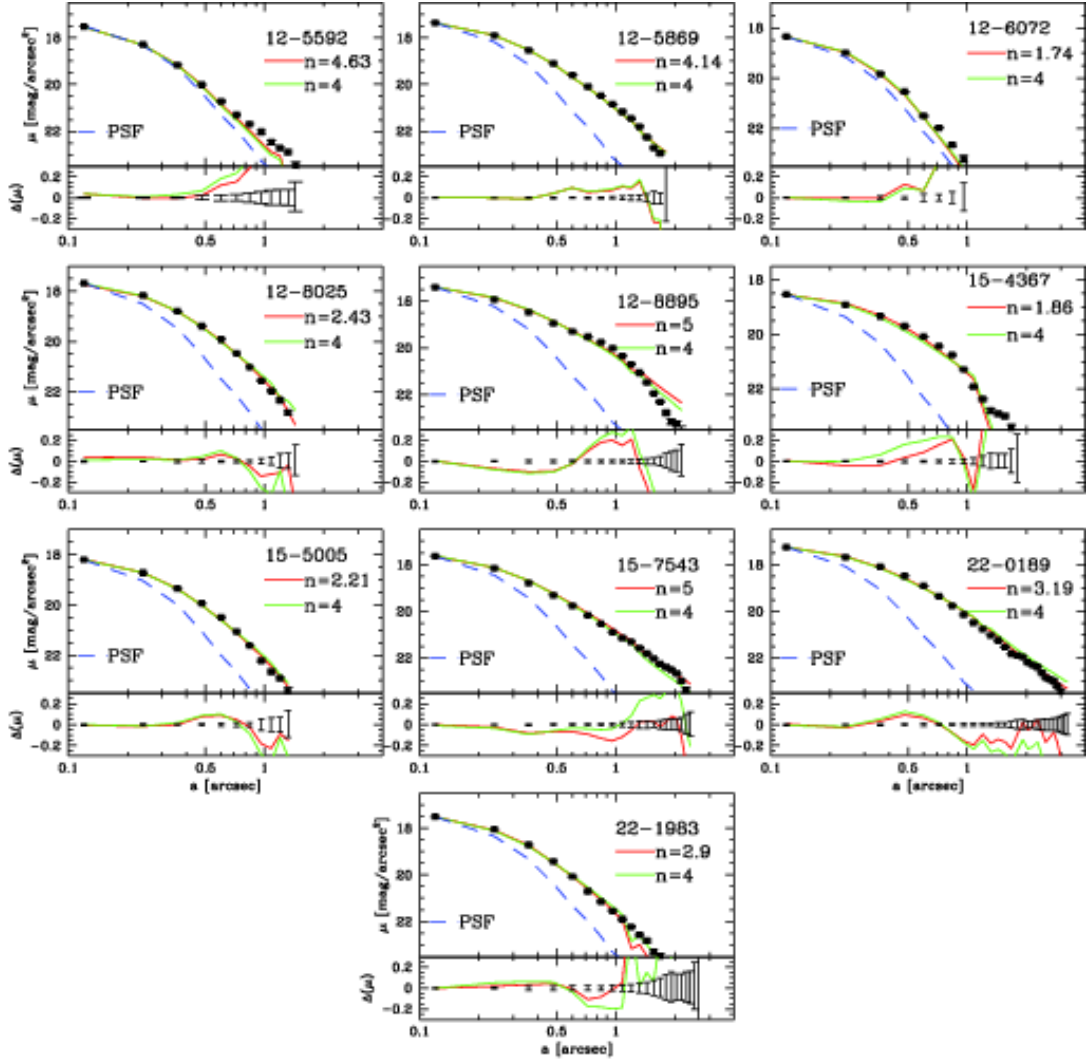}
\caption{Upper panels: major axis surface brightness profiles in the F160W band for each galaxy 
(squares), with R$^{1\slash 4}$ (green line) best-fit profile, R$^{1\slash n}$ (red line) best-fit profile, and a PSF profile (blue dashed line) overploted.  
The step used to present isophotal surface brightness corresponds to the pixel scale of our drizzled NIC3 images (0\farcs12). The limiting surface brightness in each panel presents (roughly) 5$\sigma$ limit for our observations. The lower part of each panel shows the residual differences between the data points and the model fits, 
with the $1~\sigma$ errors on the data shown for comparison.}
\label{f3}
\end{figure*}

As a consistancy check, we also determined one-dimensional (1D) azimuthally averaged radial 
surface brightness profiles for each galaxy and for the corresponding  models resulting from 
its 2D profile fits. These 1D radial profiles were extracted using the approach developed by \citet{jed87} as implemented
in IRAF \citep{tody93}.  Integrated magnitudes were determined within a series of elliptical isophotes,
the spacing of which grows with radius. We masked objects closer than $10^{\arcsec}$
before determining the surface brightness profiles of the galaxies. 
In most cases we are able to determine the profile over roughly six magnitudes of surface 
brightness and to radii of $1\farcs5$, or $\sim13$~kpc at $z = 1.5$. 
The $5\sigma$ limiting surface brightness for most of our observations is $\mu_{F160}\approx23$ mag~arcsec$^{-2}$;
the data for 12-5869 and 12-5592 reach approximately 0.3 magnitudes
deeper.  This surface brightness limit corresponds to $\mu_{r}\approx20$ mag~arcsec$^{-2}$ ($\mu_{r}\approx20.3$ mag~arcsec$^{-2}$ for 12-5869 and 12-5592) 
for a galaxy at redshift $z=1.5$ that is formed at $z_f=6$ with exponentially declining SFR and $e$-folding time $\tau=0.1$~Gyr. 
Surface brightness profiles were determined in a similar fashion for each star that served as a local measure of the PSF.  Azimuthally averaged surface brightness profiles for all of our GDDS
objects are presented in Figure~\ref{f3},  with the profiles of best-fitting 2D models 
and a PSF profile shown as solid lines and a dashed line, respectively. Figure~\ref{f3}  
confirms that all galaxies in our GDDS sample are well resolved, except for the target 12-6072 that seems only 
marginally resolved when compared to the PSF 1D profile. The profiles are smooth in nearly all cases, the exception being object 15-4367 which
shows a step at $\rm{a}=1\farcs5$.  
Careful examination of this obejct's NIC3 image revealed that it was not perfectly symmetric and harboured a weak disk. 
The best R$^{1\slash n}$ profile index of $\sim2$ confirms these findings. 
In addition, 15-4367 has a very faint neighbouring object that had to be masked out before fitting. 
These two effects produced the step in its 1D profile seen in Fig.~\ref{f3}.

\begin{deluxetable}{lcccccccccc}
\tablecolumns{7}
\tablewidth{0pc}
\tablecaption{Morphological parameters of the galaxies in the GDDS sample\label{tab3}}
\tabletypesize{\tiny}
\tablehead{
\colhead{ID} & \colhead{F160W} & \colhead{n} & \colhead{r$_e$} & \colhead{R$_{e}$}  & \colhead{$\langle\mu\rangle_e^{160W}$} & \colhead{$\langle\mu\rangle_e^{corr}$} & \colhead{$b/a$} & \colhead{$\chi^2$} \\
\colhead{}& \colhead{[mag]} &\colhead{} &  \colhead{[arcsec]} & \colhead{[kpc]} & \colhead{[mag/arcsec$^2$]}& \colhead{[mag/arcsec$^2$]} & \colhead{}& \colhead{}}
\startdata
12-5592\dotfill&21.60 $\pm$ 0.04 &    4 &     0.05 $\pm$ 0.03 &  0.4$^{+0.2}_{-0.3}$ & 17 $\pm$  1  &  14 $\pm$  1  & 0.9$^{+0.08}_{-0.4}$ & 0.205\\
&21.58 $\pm$ 0.07 &       5 $\pm$ 2 &  0.05 $\pm$ 0.05   &     0.4 $\pm$ 0.4 & 17 $\pm$ 2 & 14 $\pm$ 2 &  0.9$^{+0.1}_{0.3} $  &  0.200\\
12-5869\dotfill&20. 79$\pm$ 0.09 &    4  &    0.25 $\pm$ 0.06 &  2.1 $\pm$ 0.5 &  19.8 $\pm$ 0.5 &  16.6  $\pm$ 0.5 &  0.82 $\pm$ 0.07 &0.495\\
& 20.78 $\pm$ 0.06   &      4.1 $\pm$ 0.9 &  0.25 $\pm$ 0.04  &      2.1 $\pm$ 0.3 & 19.8 $\pm$ 0.3 & 16.7 $\pm$ 0.3 &  0.82 $\pm$ 0.07   &   0.495\\
12-6072\dotfill&22.30 $\pm$ 0.08 &        4 &     0.04 $\pm$ 0.04 &       0.3  $\pm$ 0.3 &   17 $\pm$ 3  & 14 $\pm$ 3  &      0.97$^{+0.03}_{-0.3}$   &   0.174\\
&22.31 $\pm$ 0.07   &  1.7 $\pm$ 1.2 &  0.09 $\pm$ 0.04  &  0.8 $\pm$ 0.3 & 19.1 $\pm$ 0.9 & 15.9 $\pm$ 0.9 &  0.96$^{+0.04}_{-0.2}$    &  0.172\\
12-8025\dotfill&21.05 $\pm$ 0.05&        4 &     0.25 $\pm$ 0.05 &      2.1 $\pm$ 0.4  &   20.1 $\pm$ 0.4 &   17.2 $\pm$ 0.4 &  0.79 $\pm$ 0.06   &   0.258 \\
& 21.13 $\pm$ 0.03   &  2.4 $\pm$ 0.6 &  0.24  $\pm$ 0.03 & 2.0$\pm$ 0.2 & 20.0 $\pm$ 0.3 & 17.1 $\pm$ 0.3 & 0.89 $\pm$ 0.06    &  0.240\\
12-8895\dotfill&20.6 $\pm$ 0.2  &   4  &   0.3 $\pm$  0.1 &   2.9 $\pm$ 0.7 &  20.2 $\pm$ 0.6 &    16.8 $\pm$  0.6 &  0.42 $\pm$ 0.06 & 0.418\\
& 20.44 $\pm$ 0.04   &  5.0$\pm$ 0.6   &  0.50 $\pm$ 0.05 &  4.2 $\pm$ 0.4 & 20.9 $\pm$ 0.2 & 17.5 $\pm$ 0.2 & 0.52 $\pm$ 0.04   &   0.407\\
15-4367\dotfill&21.81 $\pm$ 0.06    &     4  &    0.19 $\pm$ 0.04  &  1.6 $\pm$ 0.3 &    20.2  $\pm$ 0.4 & 16.7 $\pm$ 0.4 & 0.22 $\pm$ 0.05 &  0.248\\
&  21.91  $\pm$ 0.03  &  1.9 $\pm$ 0.3 &  0.22 $\pm$ 0.03 &  1.9 $\pm$ 0.2 & 20.6 $\pm$ 0.3 & 17.1 $\pm$ 0.3 & 0.32 $\pm$ 0.06   &   0.231\\
15-5005\dotfill&21.69 $\pm$ 0.05 &     4   &  0.17 $\pm$ 0.05   &  1.4 $\pm$ 0.4  &   19.8 $\pm$ 0.6     & 16.1 $\pm$ 0.6       &       0.74  $\pm$ 0.08&    0.247\\
& 21.73 $\pm$ 0.03    &  2.2 $\pm$ 0.6  &0.21 $\pm$ 0.03 & 1.8 $\pm$ 0.2 & 20.4 $\pm$ 0.2 & 16.6 $\pm$ 0.2 & 0.86 $\pm$ 0.08  &    0.240\\
15-7543\dotfill &   20.86 $\pm$ 0.06 &   4  &    0.40 $\pm$ 0.06 &  3.0 $\pm$ 0.5 &  20.6 $\pm$  0.4  &   17.0 $\pm$  0.4 &  0.79 $\pm$ 0.04 & 0.275\\
& 20.71 $\pm$ 0.08   & 5.0$\pm$ 0.7  & 0.48 $\pm$ 0.08 &  4.0 $\pm$ 0.7 & 21.1 $\pm$ 0.4 & 17.4 $\pm$ 0.4 & 0.78 $\pm$ 0.04   &   0.267\\
22-0189\dotfill &  20.32 $\pm$ 0.06 &   4    &  0.42 $\pm$ 0.06 &  3.6 $\pm$ 0.5 & 20.4 $\pm$  0.3   &  17.3 $\pm$ 0.3 &  0.49 $\pm$  0.04 & 0.454\\
& 20.40  $\pm$ 0.04  &      3.2 $\pm$ 0.7 &  0.37 $\pm$ 0.03 & 3.1 $\pm$ 0.3 & 20.2 $\pm$ 0.2 & 17.1 $\pm$ 0.2 &  0.50 $\pm$ 0.04  &    0.431\\
22-1983\dotfill&   21.33 $\pm$ 0.04  &  4  &    0.09 $\pm$  0.04 &  0.7 $\pm$ 0.4 & 18 $\pm$  1  &     15 $\pm$  1 &    0.2 $\pm$ 0.2 & 0.259\\
&  21.35 $\pm$ 0.02  & 2.9  $\pm$ 0.8 &  0.09 $\pm$ 0.04  & 0.7 $\pm$ 0.4    &  18 $\pm$ 1  & 15 $\pm$ 1 & 0.2 $\pm$ 0.2   &   0.242 \\
\enddata
\end{deluxetable}

\begin{deluxetable}{lccccccccc}
\tablecolumns{7}
\tablewidth{0pc}
\tablecaption{Morphological parameters of the galaxies in the MUNICS sample\label{tab4}}
\tabletypesize{\tiny}
\tablehead{
\colhead{ID}& \colhead{F160W} & \colhead{n} & \colhead{r$_e$} & \colhead{R$_{e}$}  & \colhead{$\langle\mu\rangle_e^{160W}$} & \colhead{$\langle\mu\rangle_e^{corr}$} & \colhead{$b/a$} & \colhead{$\chi^2$} \\
\colhead{}& \colhead{[mag]} &\colhead{} &  \colhead{[arcsec]} & \colhead{[kpc]} & \colhead{[mag/arcsec$^2$]}& \colhead{[mag/arcsec$^2$]} & \colhead{}& \colhead{}}
\startdata
S2F5\_109\tablenotemark{a}\dotfill&18.57 $\pm$ 0.03 &    4 &     0.66 $\pm$ 0.03 &  5.5 $\pm$ 0.2 & 19.65 $\pm$  0.08  &   17.14 $\pm$ 0.08  & 0.48 $\pm$ 0.02 & 0.320\\
S2F5\_109\tablenotemark{b}\dotfill&18.64 $\pm$ 0.03  &   4   &  0.67 $\pm$ 0.01   &    5.57 $\pm$ 0.09 & 19.77 $\pm$ 0.04 & 17.25 $\pm$ 0.04\tablenotemark{c}& 0.49 $\pm$ 0.01 &  0.532\tablenotemark{d}\\
S7F5\_254\tablenotemark{a}\dotfill&20.42 $\pm$ 0.02 & 4 & 0.36 $\pm$ 0.01 & 3.00 $\pm$ 0.08 & 20.20 $\pm$ 0.07 & 17.68 $\pm$ 0.07& 0.90 $\pm$ 0.01 & 0.265\\
S7F5\_254\tablenotemark{b}\dotfill&20.56 $\pm$ 0.03 & 4 & 0.34 $\pm$ 0.01 & 2.80 $\pm$ 0.11 &  20.20 $\pm$ 0.09 & 18.86 $\pm$ 0.09\tablenotemark{c} & 0.83 $\pm$ 0.02 & 0.402\tablenotemark{d}\\
S2F1\_357\tablenotemark{a}\dotfill& 19.80 $\pm$ 0.03 &    4  &    0.41 $\pm$ 0.02 &  3.4 $\pm$ 0.1 &  19.9 $\pm$ 0.08 &  17.08  $\pm$ 0.08 &  0.67$\pm$ 0.01 &0.312\\
S2F1\_357\tablenotemark{b}\dotfill& 19.89 $\pm$ 0.03 &      4 &  0.39 $\pm$ 0.01   &    3.28 $\pm$ 0.07 & 19.84 $\pm$ 0.06 & 17.07 $\pm$ 0.06\tablenotemark{c} & 0.66 $\pm$  0.01   &   0.440\tablenotemark{d}\\
S2F1\_389\tablenotemark{a}\dotfill& 20.99 $\pm$ 0.05& 4 & 0.23 $\pm$ 0.02  &  1.9 $\pm$ 0.2  &  19.8 $\pm$ 0.2 & 16.9 $\pm$ 0.2 & 0.93 $\pm$ 0.07 & 0.312\\
S2F1\_389\tablenotemark{b}\dotfill&21.21 $\pm$ 0.03   &  4 & 0.18 $\pm$ 0.02   &  1.54 $\pm$ 0.15 &  19.52 $\pm$ 0.24 & 16.58 $\pm$ 0.24\tablenotemark{c}  & 0.86 $\pm$ 0.03 &  0.340\tablenotemark{d}\\
S2F1\_511\tablenotemark{a}\dotfill&20.35 $\pm$ 0.05 &        4 &  0.22 $\pm$ 0.02 & 1.9 $\pm$ 0.2 & 19.1 $\pm$ 0.2 &  16.2 $\pm$ 0.2 & 0.59 $\pm$ 0.05 & 0.269\\
S2F1\_511\tablenotemark{b}\dotfill& 20.43 $\pm$ 0.03  & 4& 0.23 $\pm$ 0.01 & 1.91 $\pm$ 0.07 & 19.21 $\pm$ 0.09 & 16.33 $\pm$ 0.09\tablenotemark{c} &0.59 $\pm$ 0.01&0.343\tablenotemark{d}\\
S2F1\_142\tablenotemark{a}\dotfill& 20.06 $\pm$ 0.03  &   4  &   0.62 $\pm$  0.03 &   5.2 $\pm$ 0.2 &  21.02 $\pm$ 0.09 &    18.05 $\pm$  0.09 &  0.79 $\pm$ 0.02 & 0.309\\
S2F1\_142\tablenotemark{b}\dotfill& 19.95 $\pm$ 0.03 &      4  &  0.35 $\pm$ 0.01&  2.95 $\pm$ 0.7 &  19.67 $\pm$ 0.06 & 16.70 $\pm$ 0.06\tablenotemark{c} & 0.73 $\pm$ 0.01 &0.915\tablenotemark{d}\\
S7F5\_045\tablenotemark{a}\dotfill& 19.73 $\pm$ 0.02 & 4 & 1.00 $\pm$ 0.02 & 8.5 $\pm$ 0.2 & 21.73 $\pm$ 0.05 & 18.72 $\pm$ 0.05& 0.70 $\pm$ 0.02 & 0.389\\
S7F5\_045\tablenotemark{b}\dotfill& 19.61 $\pm$ 0.03 & 4 & 1.13 $\pm$ 0.04 & 9.53 $\pm$ 0.33 & 21.87 $\pm$ 0.09 & 18.10 $\pm$ 0.09\tablenotemark{c} & 0.69 $\pm$ 0.01 & 0.394\tablenotemark{d} \\
S2F1\_633\tablenotemark{a}\dotfill&20.98 $\pm$ 0.03    &     4  & 0.31 $\pm$ 0.02 &  2.6 $\pm$ 0.1 &  20.4 $\pm$ 0.1  & 17.4 $\pm$ 0.1 & 0.56 $\pm$ 0.02 &0.301\\
S2F1\_633\tablenotemark{b}\dotfill& 20.36 $\pm$ 0.03   & 4&  0.26 $\pm$ 0.01& 2.23 $\pm$ 0.07 &19.46 $\pm$ 0.08  &16.42 $\pm$ 0.08\tablenotemark{c} & 0.53 $\pm$ 0.01&1.258\tablenotemark{d} \\
S2F1\_443\tablenotemark{a}\dotfill& 20.96 $\pm$ 0.08 & 4 & 0.81 $\pm$ 0.06 & 6.9 $\pm$ 0.5 & 22.5 $\pm$ 0.2 & 19.0 $\pm$ 0.2 &0.81 $\pm$ 0.05& 0.252 \\
S2F1\_443\tablenotemark{b}\dotfill& 20.30 $\pm$ 0.03 & 4 & 0.72 $\pm$ 0.03 & 6.13 $\pm$ 0.24 & 21.6 $\pm$ 0.1 & 18.1 $\pm$ 0.1\tablenotemark{c} &  0.76 $\pm$ 0.02 & 0.676\tablenotemark{d}\\
\enddata
\tablenotetext{a}{\ our best fit parameters for MUNICS sample} 
\tablenotetext{b}{\ best-fit model from \citet{lon07}}
\tablenotetext{c}{\ mean effective surface brightness correction includes K correction and $(1+z)^4$ dimming factor}
\tablenotetext{d}{\ $\chi^2$ of our fit with the parameters from \citet{lon07}}
\end{deluxetable}

In order to estimate the errors on parameters obtained by our 2D and 1D fitting procedures, we 
undertook a series of Monte Carlo (MC) simulations which incorporated all the sources
of systematic and random errors we were able to identify.  We constructed a set of galaxy images from
our best-fit model for each galaxy and convolved these with a range of PSFs (i.e., PSFs obtained 
from different stars) and added these to the background images. We dithered the position image 
about the central value to explore the importance of binning, and used RMS images to construct
2D arrays of random numbers to capture poisson noise and structure in the sky background. 
Each image constructed in this way went through the same fitting procedure as 
the real galaxy image from our sample. The standard deviations of resulting parameters are 
shown as the error estimates reported in Table~\ref{tab3}. The reduced $\chi^2$ values for 
the best fits to the MC simulations are of the order of unity and larger then reduced 
$\chi^2$ of the best fits to the data, which makes our error estimates very conservative.

\subsection{K-corrections and cosmological dimming}  

Our analysis requires comparison between the properties of our $1.2 < z < 2$ samples
observed at 1.6~$\mu$m ($H$-band) to those of present-day galaxies observed at visible
wavelengths. In order to make a proper comparison, we need to transform
the various data sets to a common bandpass and apply a K-correction.
We computed appropriate spectral energy distributions (SEDs) using
PEGASE-HR spectral synthesis models \citep{lb04}. The
model that we used is based on the \citet{bal03} initial mass function (IMF), solar metallicity, and an exponentially declining
star formation rate with a time scale of $\tau = 0.1$~Gyr, very similar to a single burst.
The typical ages of GDDS and MUNICS passive galaxies at $1.2 < z < 2$ are $3-4$~Gyr (\citetalias{mcc04}, \citealp{lon05}) and
we used a 4~Gyr model to approximate their SED. It is important to emphasize that the correction needed to reduce our $H$-band
data to rest-frame SDSS-$r$ is remarkably insensitive to 
SED shape since
redshifted $H$-band closely matches rest-frame SDSS-$r$ at $z\sim1.5$.
The photometry for the two samples is listed in Tables~\ref{tab3} and \ref{tab4}.
Cosmological surface brightness dimming will reduce the observed surface brightness
and these must be corrected by $(1 + z)^4$ to transform them to the rest-frame.

\section{Results}\label{res}

\subsection{Morphologies of Passive Galaxies at $z>1.3$}

All of the objects in our NICMOS F160W sample (shown in
Figures~\ref{f1} and \ref{f2}) have compact morphologies and none 
show obvious evidence of interactions, such as double nuclei or disturbed isophotes 
at bright levels. The star-forming massive galaxies drawn from the GDDS sample, 
by contrast, exhibit a wide range of disturbed morphologies as shown in \citetalias{abr07}.
The intermediate age and composite population systems primarily have disk morphologies, 
while the passive galaxies at $z<1.3$ discussed in \citetalias{abr07} exhibit a preponderance 
for compact and regular morphologies.  Six of the 10 GDDS galaxies in the present sample
appear to be early types with R$^{1\slash n}$ profile index $n>2.5$ (Table~\ref{tab3}), while the four potential disk systems in our $z>1.3$ passive sample appear 
to have prominent bulges. Thus $60\%$ of our GDDS sample defined by spectral properties 
have pure early type morphologies, 
and this fraction rises to $90\%$ when the prominent bulges with very weak disks are also taken into account as early type object.
To a first approximation, our NICMOS Camera 3 images extend the correlation between spectra indicative of old stellar populations and 
compact early-type morphologies from $z \sim 1.3$ to $z \sim 2$. This is not surprising 
given previous indications in this direction from smaller samples \citep[e.g.,][]{cim04}. 

The correlation between color and morphological type is not as strong for the red 
galaxies, as a number of studies have shown. At redshifts near unity, red $R-K$ 
or $I-K$ selected samples contain roughly as many disk as early-type galaxies 
\citep[e.g.,][etc.]{mou04,yan03}. At higher redshifts red selected 
samples also show a mix of morphologies, as shown for the $z\sim1.5$ range in 
\citetalias{abr07} and at  $z>2$ by \citet{la05}, \citet{sto04}, and others. 

\subsection{Surface Brightness Profiles \& Sizes}

Azimuthally averaged surface brightness profiles presented in 
Figure~\ref{f3} confirm that six of our 10 GDDS galaxies are well-fit by R$^{1\slash 4}$ profiles.
The effective radii for these six objects range from as small as $0\farcs05$ to as large as $0\farcs42$, or
from 0.4 to 3.6~kpc. The median effective radius is $0\farcs26$ or 2.2~kpc.
As Figure~\ref{f1} shows, for the most part the 2-D models fit the data well and the residuals are not
significantly greater than the sky noise. In 12-8895 and 12-5869 there appear
to be some non-axisymmetric structures within the central one arcsecond, while
in 12-6072 the model is too peaked.  Four of our 10 GDDS galaxies are clearly better 
fit by R$^{1\slash n}$ profiles with indicies near 2, rather than the R$^{1\slash 4}$~law. These are: 12-6072, 12-8025, 
15-4367 and 15-5005. As can be seen 
in Figure~\ref{f3}  the significance with which the R$^{1\slash 4}$~law fit is rejected 
in these objects
is low except in the case of 12-8025 where the outer isophotes depart strongly from 
the R$^{1\slash 4}$~law profile.

The effective radii of the GDDS galaxies are smaller than those of present-day cluster
ellipticals and early-type field galaxies. The median effective radius for low redshift cluster ellipticals is $\sim 4$~kpc 
\citep{jor95,schombert86}, and the field early type galaxies at $z\sim 0.5$
from the CFRS \citep{sch99} have a fairly similar median size. The hosts of luminous radio galaxies
at $z \sim 0.8 - 1$ studied by \citet{zi03} probably represent the most massive end
of the field \& group early type populations at these redshifts. Their sizes are also
similar to the lower redshift samples and larger than the GDDS elliptical galaxies that have median effective radius of 2.2~kpc. 
In contrast, the distant red galaxies (DRGs), defined by their $J-K$ colors, at $2 < z < 3$ have a median 
effective radius of 1.4~kpc \citep{to07}, somewhat smaller than the passive GDDS galaxies in
our sample at $z \sim 1.7$. 
 
The sizes of the GDDS passive galaxies appear to support a fairly strong evolution
in scale length among the early type galaxies in the $1 < z < 3$ interval. 
A mundane potential explanation for this result is that the under-sampling of the 
NIC3 PSF data has led to unreliable fits. We can
rule out this hypothesis on the basis of three tests. Firstly, we have
re-fitted the six galaxies with more finely sampled NIC2 data
from the \citet{lon07} sample, and we recover very similar fits (see Table ~\ref{tab4}). 
These fits are shown in Figure~\ref{f4} using dashed lines to join the values of points obtained by \citet{lon07}
to those obtained by us. Secondly, we have undertaken detailed MC
simulations (used to set our error bars in Figure~\ref{f4}) based on
generating idealized over-sampled images which are randomly displaced
by sub-pixel shifts before being binned to NIC3 resolution and re-fitted.
Lastly, two of our objects - 12-5592 and 22-1983 - were observed in the F814W band with ACS on HST.
The sizes that we measure for these galaxies, albeit at shorter rest-frame
wavelengths, are in good agreement with the sizes derived from our NIC3 data.
Thus we are confident that our size determinations are robust.

The strong correation between mass and size,
as measured by the effective radius, makes comparisons between the average or median
properties of different samples imprecise measures of evolution. The lower redshift samples
\citep[$z<1$][]{jor95,schombert86,sch99} cover a broad range of the parent luminosity functions while
the higher redshift objects ($1<z<3$),
including the DRGs, the GDDS and MUNICS samples (\citealp{to07}, \citetalias{gla04}, \citealp{lon05}), sample the 
high mass end of the galaxy population and thus are biased to large values in their median sizes.  
This further strengthens the conclusion that there is strong evolution in the characteristic
sizes of early type galaxies above $z \sim 1$. The evolution in galaxy sizes can be further 
quantified by examining the size-mass correlation and its evolution, as is discussed in section~\ref{sizemass}.

\subsection{The Kormendy Relation to $z = 2$}

\begin{figure*}[htp]
\epsscale{1.5}\plotone{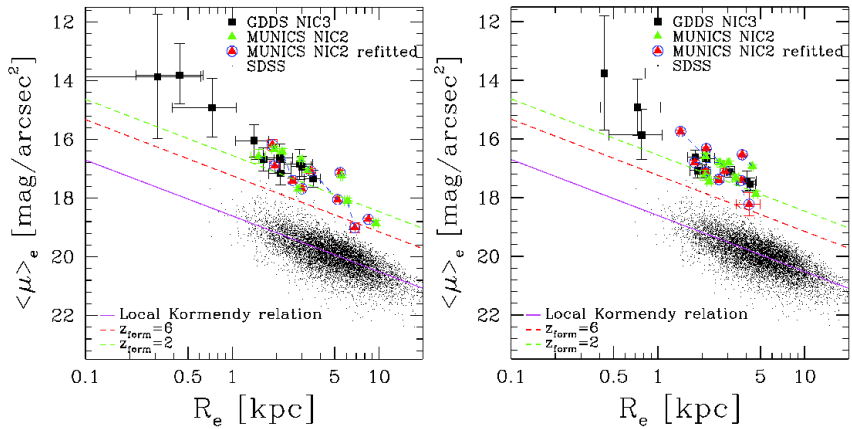}
\caption{Mean rest frame Gunn-$r$ surface brightness within effective radius R$_e$ as a function 
of R$_e$ (Kormendy relation) for objects at redshifts $1.2<z<1.9$ (GDDS and MUNICS samples) and 
for the sample of local galaxies \citep[SDSS,]{ber03}. 
The solid line is the best-fit relation to the SDSS objects. The dashed lines represent the expected luminosity evolution of the 
local (SDSS, solid line) relation at $z=1.5$ for galaxies formed at $z_{form}=6,2$ 
with exponentially declining SFR and $e$-folding time $\tau=0.1$~Gyr. Different symbols 
correspond to different samples, and circled triangles denote  
re-fitted MUNICS sample. Left panel shows R$^{1\slash 4}$~profile parameters for the galaxies from the GDDS and MUNICS samples, while the right one
shows their best-fit R$^{1\slash n}$~profile parameters.}     
\label{f4}
\end{figure*}

In Figure~\ref{f4} we present the rest-frame $r$-band Kormendy 
relation, $\langle\mu\rangle_e$ vs. R$_e$, for the GDDS and MUNICS samples. As noted earlier,
our construction of this diagram is particularly robust because our
observed $H$-band observations match rest-frame $r$-band 
at $z=1.5$ and hence there is negligible residual K-correction uncertainty. 
We have not applied any evolutionary corrections to the observed surface brightness values.
Figure~\ref{f4} includes the corresponding distribution for present-day early-type 
galaxies from the SDSS \citep{ber03}.

Figure~\ref{f4} shows that the tightness and
slope of the Kormendy relation in the GDDS + MUNICS sample is
similar to that defined by the local relation. There is a hint that
the high-redshift slope may be slightly 
steeper than the local value, but the difference is not significant. 
While the high-redshift ellipticals fall along a tight Kormendy relation, the relationship itself is 
offset to higher surface brightness from the low-redshift reference sample. 
The simplest explanation for this is that is that we are seeing galaxies nearer 
to their epoch of formation, when they are brighter, and thus the Kormendy relation is shifted 
upwards. This evolutionary effect cannot fully explain the evolution in the Kormendy relation.
The offset in surface brightness compared to the
$z \sim 0$ sample is too large ($\sim2.5$ mag) to be explained by pure luminosity evolution of stellar 
populations unless the redshift of formation is very recent ($z_{\rm form}\lesssim2$),
which is inconsistent with both their colors and spectra \citepalias[see][]{mcc04} which 
argue that these are old systems with $z_{\rm form}\gtrsim4$. In the latter case,
the maximum dimming allowed is 1 to 1.5~mag, depending on the selected IMF and the star formation history.  In addition, we also see from this figure that, 
in spite of their large masses, typical high-z ellipticals are
substantially smaller than their local counterparts.  In contrast to the median effective radius for the GDDS sample of 2.2~kpc, early-type galaxies in the SDSS sample presented in 
Fig.~\ref{f4} span the range of effective radii with the median value of 4.9~kpc. Finally,
we see that three out of ten galaxies in the GDDS sample are
`ultra-compact' (R$_e<1$ kpc), and thus are of much higher stellar density. 
\citet{cim08} found a similar fraction from ACS imaging and estimate that the number density of 
comparably dense objects at $z=0$ is up to $10^4$ times lower than at $z=1.5$.  
In contrast, in the MUNICS sample of elliptical galaxies ($1.2<z<1.7$) no `ultra-compact' objects are found.
As we will discuss in the following section, our findings lead us to also conclude that 
strong size evolution (a factor of 2 or more) is the additional ingredient needed to explain 
the shift in the Kormendy relation. 

\subsection{The Mass-Size Relation and the Stelar Mass Kormendy Relation}\label{sizemass}

As the previous section illustrates, a proper comparison between galaxy samples at high
and low redshifts nearly always entails corrections for luminosity evolution. We can, however, improve on the standard procedure of
using simple models of luminosity evolution by using multi-color
SED data to fit stellar  population models and derive stellar masses for the galaxies in question
(this was done and described in \citetalias{gla04} for the GDDS sample). We
then recast the data into a new `stellar mass Kormendy relation' which allows a more fundamental
comparison. By doing this we are using the complete set of information (the colors) to
measure and remove the luminosity evolution. A further advantage to the use of stellar mass is that
it allows us to compare optical and near-IR samples and plot them on the same diagram. A possible disadvantage is that 
we rely heavily on the mapping from light to stellar mass given by our spectral synthesis modeling, which,
in turn, depends on the correctness of our assumptions.  So for
example derived masses would be in error if the assumed IMF is
evolving rather than static.

We consider two projections of the
structural evolution that minimize the impact of luminosity and spectral evolution. The first
is the size-mass relation, while the second is the relation between stellar mass 
density and size, which we will refer to as the stellar mass Kormendy relation. 
In deriving the stellar mass {\em density} we assume
that the F160W light traces the stellar mass. 

\begin{figure*}[htp]
\epsscale{2}\plotone{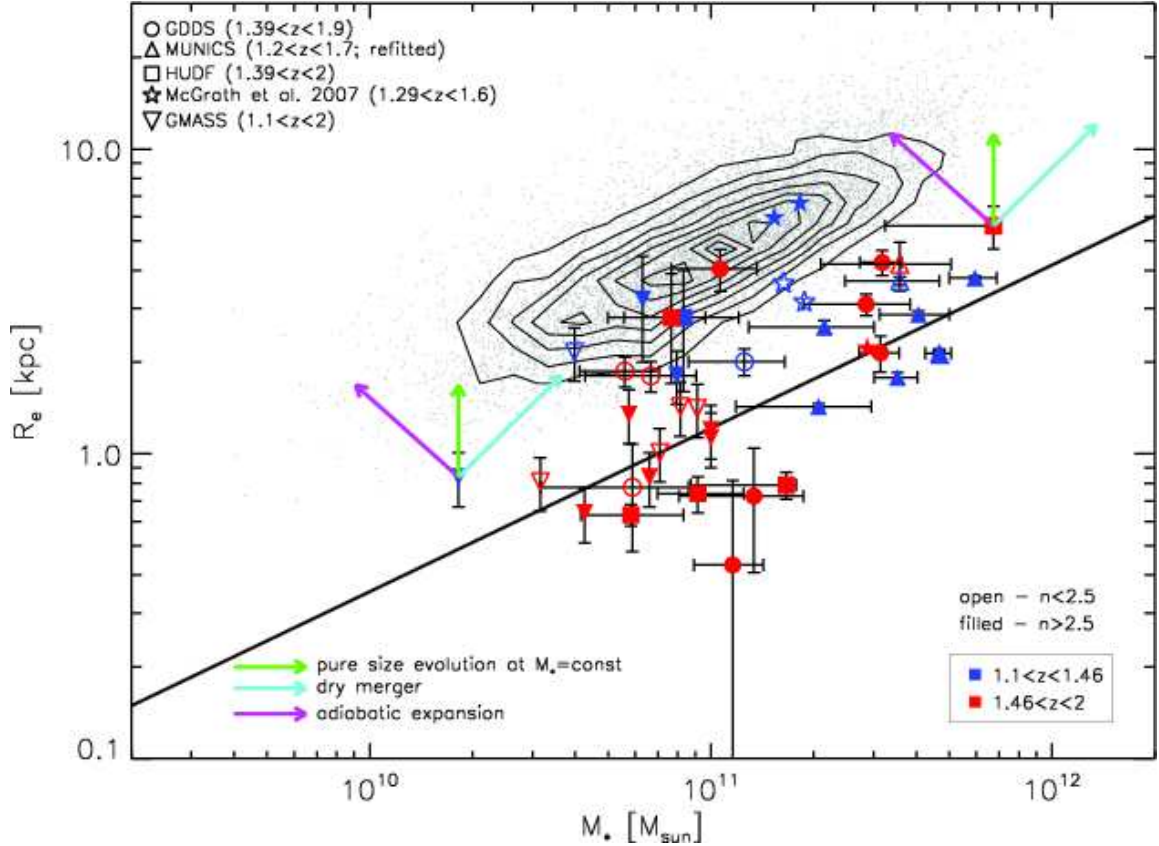}
\caption{Effective radius $\rm{R}_e$ as a function of stellar mass for five samples of early-type galaxies in the redshift range $1.1<z<2$. Points are color-coded by two redshift ranges (red = $z>1.46$, blue = $z<1.46$).
Different symbols correspond to different surveys, with triangles denoting re-fitted object from the MUNICS sample (as in Fig.~\ref{f4}). The
size-mass relation for local early-type galaxies in the SDSS is presented with sizes taken from \citet{ber03}, and matched with 
masses calculated following \citet{bal08} (black points). Contours represent linearly spaced regions of constant density of galaxies in size-mass parameter space. The solid line is the best-fit relation 
to the data points at redshifts $1.1<z<2$.  Three arrows denote the effects that 1:1 dry merger \citep{bk06}, adiabatic expansion with $50\%$ mass loss, and pure size evolution at constant stellar mass would have on the positions of both the least and the most massive galaxy. See text for details.}
\label{f5}
\end{figure*}

In Figure~\ref{f5} we plot the size-mass relation for our sample. To enhance the usefulness 
of this figure, we augmented our GDDS and MUNICS data using
published measurements obtained for passive galaxies in the redshift range $1.1<z<2.0$
taken from from two surveys in the HUDF \citep{dad05b,mar06},
a survey of six galaxies with dominant old stellar population in the fields of
radio-loud quasars \citep{mcg07a,mcg07b},
and GMASS \citep{cim08}.  While \citet{mcg07a} use NIC3 F160W observations for their morphological analysis, 
GMASS \citep{cim08} and HUDF \citep{dad05b} effective radii were measured by fitting ACS F850LP ($z$ band) galaxy images.
We corrected all of the stellar mass determinations to a common IMF, 
using \citet{bal03} IMF, according to the relations given in \citet{cim08} and \citetalias{gla04}.
Finally, to place our data in a broader context, Figure~\ref{f5} also shows the size-mass
relationship for local early-type galaxies in the SDSS \citep{ber03}. We
recomputed the stellar masses for the \citet{ber03} SDSS sample using the same prescription applied
to the GDDS sample (\citealp{bal08}; \citetalias{gla04}). The derived masses are in good agreement with those of \cite{kau03}. 
The size-mass relationship for early-type galaxies
shown in Figure~\ref{f5} shows a number of interesting
features, the most striking of which is that the high-redshift
and low-redshift populations show relatively little overlap. In fact, they seem to describe nearly
independent loci in size-mass parameter space, with similar slopes, but
with galaxies at $z=1-2$ systematically smaller, at a fixed mass, than galaxies at $z=0$.
The error bars on individual data points are rather large, but taken as a whole, only $\sim25\%$ of high
redshift early-type galaxies lie in the region of size-mass space occupied by low-redshift systems.

The size-mass relationship of elliptical galaxies at $z\sim0$ is well described by a
power law with the same exponent  ($\sim0.5$) as for the early-types at $z\sim1.5$.
Galaxies with stellar masses of $8 \times 10^{10}$~M$_{\odot}$, comparable to M$^*$ today,
are approximately three times smaller at $z \sim1.5$ than their apparent counterparts today.
The number density of  compact galaxies with R$_e<1$~kpc (`red nuggets') in the redshift range $1.1<z<2$
is $2\times10^{-5}$~Mpc$^{-3}$. In contrast, number density of these objects in the SDSS sample \citep{ber03} 
is $3\times10^{-8}$~Mpc$^{-3}$, three orders of magnitude lower than that for the higher redshift objects. The
`red nuggets' in two samples are different with respect to mass, too - the median of GDDS compact galaxies 
mass is $10^{11}$~M$_{\sun}$, while the objects of the same compactness  in the local Universe have masses 
with ten times lower median (i.e.,~$10^{10}$~M$_{\sun}$). The passive galaxy population at $1.1 < z < 2$ span a similar range in stellar mass as
galaxies today ($2 \times 10^{10} - 6\times 10^{11}$~M$_{\odot}$) so, at least at the high mass end, the bulk of the evolution 
from $z \sim 2$ to $z \sim 0$ appears to be in size rather than mass. 

In Figure~\ref{f6} we plot the projected stellar 
mass density within a radius equal to R$_e$ (i.e, $\rho_e=3M_{*}(\rm{R}<\rm{R}_e)\slash (4\pi$R$_e^3$)) versus R$_e$ - the
stellar mass Kormendy relation. This projection shows the evolution in the
structural properties of the passive early-type galaxies very clearly. The $z > 1.1$ galaxies
are offset to smaller radii and dramatically higher projected surface mass densities
compared to massive early-type galaxies today.  
Compact objects in the local SDSS sample appear less dense since they are less massive than high redshift objects with the same size. 
In the density space populated by red nuggets at higher redshifts ($\rho_e>10^{10}$~M$_{\sun}$~kpc$^{-3}$),
there are no galaxies in the SDSS sample, implying that number density of these objects at $z=0$ is $\lesssim4\times10^{-9}$~Mpc$^{-3}$.

In both figures \ref{f5} and \ref{f6} we have color coded the symbols according to redshift into
two sub-samples: $1.1 < z < 1.46$ and $1.46 < z < 2$. This splits the sample into two 
equal time intervals of duration 1.1~Gyr and nearly equal sample sizes.
There is a signifcant diffence in the size distributions in the two sub-samples. In the
lower redshift sub-sample 6\slash 18 galaxies, or $\sim 33$\% of the sample, fall within the range
of the local sample, while in the high redshift sample, only 4\slash 25, or $\sim 17$\% of the galaxies 
fall within the locus of the local systems. Thus it appears that the strongest evolution in size
is occuring in the $1 < z < 1.5$ interval, although as we will describe in the
next section, the heterogenous nature of the data does not allow us to conclude
this with much confidence. A number of other studies \citep[e.g.,][]{treu05} show
that $z \sim 1$ early type galaxies have normal sizes and mass densities. 

\section{Discussion}\label{disc}

\begin{figure*}[htp]
\epsscale{2}\plotone{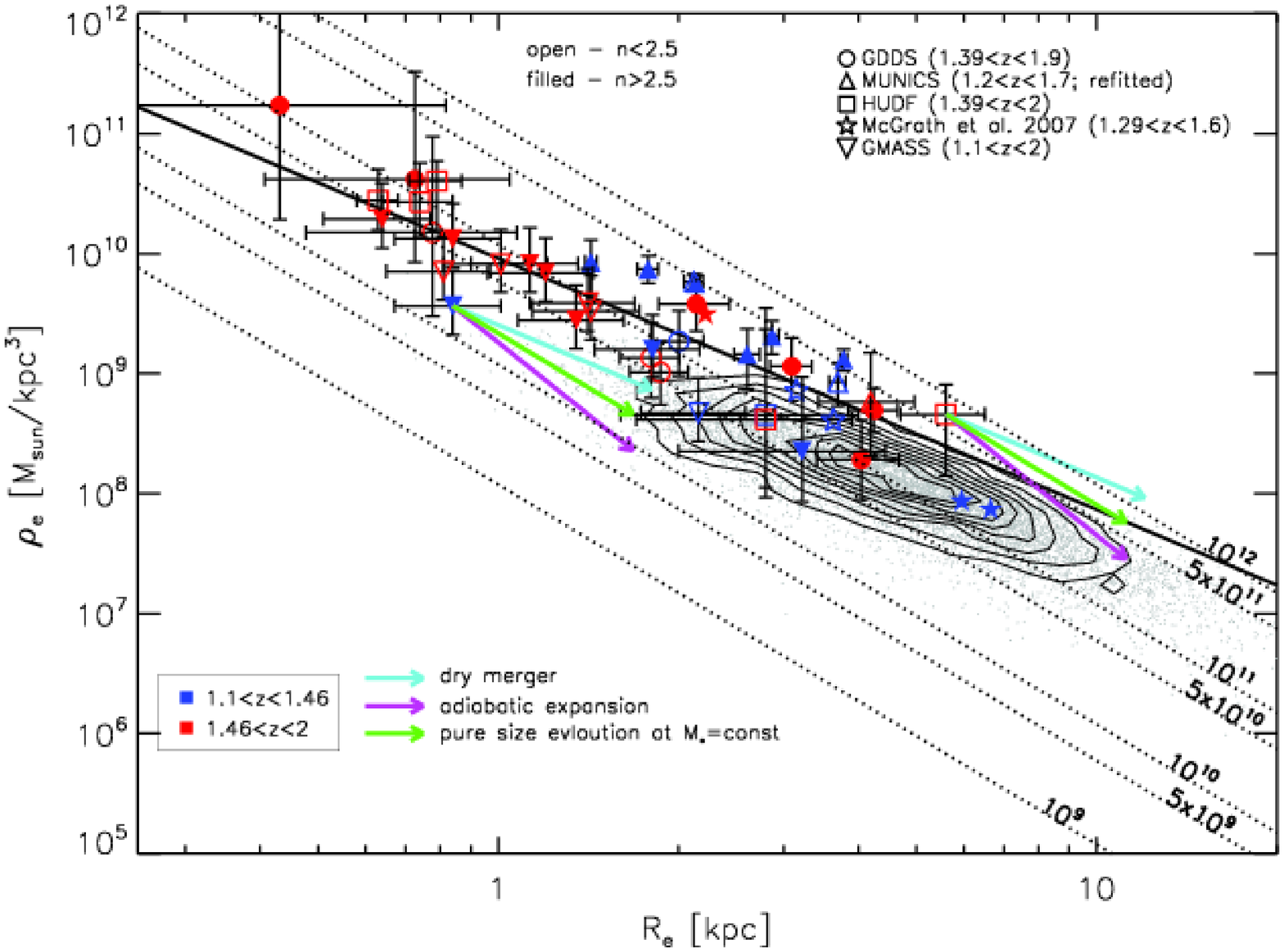}
\caption{Stellar mass density within the effective radius $\rm{R}_e$ as a function of $\rm{R}_e$ (the "stellar mass Kormendy relation'') for five samples of 
early-type galaxies in the redshift range $1.1<z<2$. Symbols are as in Fig.~\ref{f5}. The local 
sample of SDSS galaxies is presented with both points and overlaid contours that denote 
linearly spaced regions of constant density of galaxies in this parameter space. Dotted lines present the loci of constant total stellar mass, noted on each line in units of M$_{\sun}$. 
The solid line is the best-fit relation to the data points at redshifts $1.1<z<2$. Three arrows denote the effects that 1:1 dry merger \citep{bk06}, adiabatic expansion with $50\%$ mass loss, and pure size evolution at constant stellar mass. See text for details.}
\label{f6}
\end{figure*}

The key result of this paper are that the sizes and projected mass densities of early-type
passively evolving galaxies have changed very significantly since $z \sim 2$. A number of other 
studies, noted above, have reached similar conclusions in samples with higher and overlaping
redshift intervals. Our analysis has removed much of the uncertainty associated with 
evolutionary corrections  in luminosity and spectral shape by dealing with the mass 
density rather than surface brightness. 

There are a number of potential explanations for the dramatic evolution in the sizes and 
densites of the passive galaxies. If the compact massive galaxies at $z \sim 2$ are to 
evolve into massive elliptical galaxies at $z \sim 0$ they must grow by a factor of 
$2-3$ in size. The two most plausible paths to this evolution are injection of energy into, or the
loss of mass from, the central regions.  One possibility is that mergers input energy into the
stellar systems and increase their equilbrium sizes. The quiescient spectra of galaxies
in the same stellar mass range at $1 < z < 1.5$ suggest that any such merger be ``dry'' and produce 
little star formation and related activity. Dry mergers have been identified as a 
likely evolutionary path for
the compact massive galaxies at $z > 2$ discussed recently by \cite{van08}.
The large stellar masses of the compact passive galaxies at $z < 2$ suggest that equal mass 
mergers cannot be ubiquitous at later epochs.  In Figures~\ref{f5} and \ref{f6} we show 
vectors that approximate the impact of an equal mass merger, based on the simulations performed by  \citet{bk06}. Galaxies become both 
larger and more massive and move primarily along the mass-radius and mass-Kormendy 
relations rather than normal to them.  This problem makes this explanation for size
evolution unsatisfactory. While there is good evidence for an increase of roughly 
a factor of two in the total stellar mass density in red sequence galaxies since $z \sim 1.3$, 
this appears to be in the form of new galaxies appearing on the red sequence rather 
than mass growth in previously passive systems \citep[e.g.,][]{fab07,bell04}.
One could perhaps appeal to many minor mergers to puff up a galaxy's size,  but
they would have to all be dry to keep a galaxy on the red sequence and numerous
enough to have a significant effect, which seems somewhat contrived.

It has been pointed out to us (N. Murray, private communication)
that adiabatic expansion is an interesting
alternative to dry merging for increasing the size of galaxies. This process
has long been familiar to stellar dynamicists \citep{hi80} and
been verified by numerical simulation \citep[e.g.,][]{bk07}.
The process has also been used to model the influence of strong stellar
winds in conditioning the Galactic globular cluster distribution \citep{zh02}. In the present context,
the potential for adiabatic expansion to explain the existence of massive small ellipticals at 
high redshift is developed in a paper by \citet[hereafter MQT]{mu08}. To motivate the present discussion, a basic version of
the some of the key theoretical ideas in the latter paper, kindly communicated to us in advance 
of publication by the authors, will be applied to the GDDS
sample here.

Adiabatic expansion will occur in any relaxed system that is losing
mass. As mass is lost the potential becomes shallower,
so the system expands in order to relax into a new stable equilibrium.  The amount that a system expands
depends on both the extent and speed of the mass loss \citep[see][for details]{zh02}. In general, if a fraction $\frac{\Delta m}{m} = (m_{\rm initial} - m_{\rm final})\slash m_{\rm initial}$ 
of the total
mass is lost on a dynamical timescale (or longer), the size of the
system increases by a factor of approximately $1 \over {1-\frac{\Delta m}{m}}$. If the
mass is lost more quickly than the dynamical timescale, then the
expansion of the system will be larger than this estimate. It is trivial to show that as the system loses
mass the dynamical timescale increases in proportion to $1\over{(1-\frac{\Delta m}{m})^2}$ while the escape 
velocity decreases as $1-\frac{\Delta m}{m}$, so there are at least
two sources of positive feedback leading to further increase the size
as the system evolves. Of course, in the extreme case where a significant fraction of the 
total mass is lost on a short timescale, the system may become unbound.

What processes might lead to mass loss in elliptical galaxies?
The obvious candidate is stellar winds from sites of active
star formation. However, the early-type galaxies being
studied here are relatively red and spectroscopically
passive, so winds from young stellar populations are
unlikely candidates for mass loss. An interesting alternative
is mass loss from evolved A and F-type stars, and
we have explored this ideas using the following toy model. We model a galaxy
as an instantaneous burst with a solar-metallicity
stellar population whose main sequence lifetime (as a function of mass) is that given in Table
5.2 of Binney \& Merrifield (1998). We assume that after leaving the main sequence
all stars more massive than 8 solar masses wind up as stellar remnants of 1.5 solar
mass, and that all stars less than 8 solar masses wind up as remnants with 0.6 solar mass.
We also assume that mass loss from stars is never recycled into future star formation
and it ouflows far out into the galaxy's potential well, or is lost completely.

\begin{figure*}[htp]
\epsscale{1}\plotone{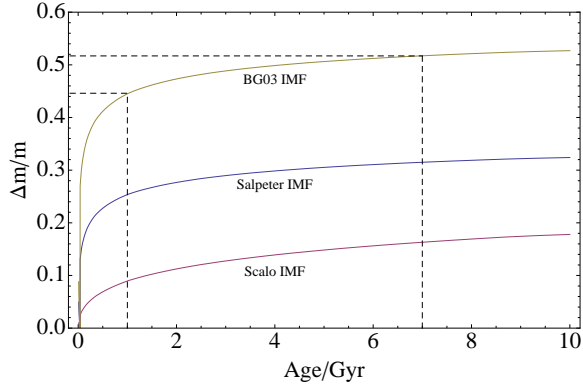}
\caption{The mass loss fraction $\frac{\Delta m}{m}$ as a function of population age in~Gyr,
for the simple model described in the text. We assume
an instantaneous burst of star formation and show $\frac{\Delta m}{m}$
as a function of time with three initial
mass functions. As expected, the total mass lost is a strong
function of the fraction of stars at the high mass end of the IMF.
The relative mass loss is small in the age range $1-7$~Gyr (dashed lines).
See text for details.}
\label{f7}
\end{figure*}

In this case, $\frac{\Delta m}{m}$ as a function of time takes on the form shown in
Figure~\ref{f7} for three initial mass functions (Salpeter IMF, Scalo IMF,
and the Baldry \& Glazebrook (2003) IMF).
Our toy model suggests that  $\frac{\Delta m}{m}$ rises sharply with time until ages of around 2~Gyr,
at which point $\frac{\Delta m(t)}{m(t)}$ flattens out, peaking at around 30\% for the
Salpeter IMF, and at 50\% for
the top-heavy Baldry \& Glazebrook (2003) IMF. Thus
the degree of mass loss from a very top-heavy IMF could explain
the size growth.
This is shown by the arrows in Figures~\ref{f5} and~\ref{f6}, which 
show the
effects that 1:1 dry merger \citep[cyan arrow]{bk06}, adiabatic expansion with $50\%$ mass loss (magenta
arrow), and pure size evolution at constant stellar mass (green arrow) have on the positions of both the least and the 
most massive galaxies in our sample. 
However, the timescale over which this
occurs poses a huge challenge for explaining the size growth entirely by
adiabatic expansion. In this paper we study the size distribution of the population 
at a time when their the stellar populations
are already rather old (see Tables~\ref{tab1} and~\ref{tab2} and discussion in
\citetalias{mcc04}) so over the redshift range
being probed the galaxies are old enough that the mass loss curves in
Figure~\ref{f7} are already nearly flat.
Another constraint on the importance of adiabatic
expansion is that is does not explain the steady factor
of (at least) three
growth in the stellar mass density locked up in massive
galaxies over the redshift range $1<z<2$
reported in \citetalias{mcc04} and in other surveys, \citep[e.g.,][]{dic03,ru06}, especially on the red sequence \citepalias{abr07}.
As the typical mass does not appear to evolve (Fig.~\ref{f5})
this primarily seems to be an evolution in number.

In spite of the problems noted above,
adiabatic expansion does appear
attractive because it
moves the high-redshift distribution shown
in Figure~\ref{f5} 
in the right direction to match
the low-redshift distribution shown in the
figure. This is not the case with 
equal-mass dry mergers, which, as shown by the cyan arrows in
the figure, and as noted by previous authors \citep{bk06},
drive evolution along the Kormendy relation rather
than displacing the relation itself.  While a top-heavy IMF loses
enough mass to grow the galaxies by the required factor
of two over their complete lifetime, the main problem with the adiabatic expansion
model is that to explain our observations that mass loss would have to occur over the 
age range of $1-7$~Gyr, 
over which Fig.~\ref{f7} shows only a $5-10~\%$ effect.  
Ages of the GDDS galaxies are taken from
Paper IV, and it is worthwhile to
consider whether we might have
significantly over-estimated the ages
of the galaxies in that paper. We think this unlikely
for two reasons. Firstly,  because broad-band
color-based ages for these galaxies seem consistent with ages
inferred from spectra
of these systems, which often exhibit photospheric
features from old stars. Secondly, because changing
to a more top-heavy IMF than the Salpeter IMF
used in \citetalias{mcc04} would not result
in systematically younger ages. In fact
the reverse is true, since 
a more top-heavy IMF would tend to produce synthetic spectra which are bluer for a given star formation history at a given age. So to match the observed colors, any fitting routine would compensate by deriving {\em older} ages for the best fit. Quantitatively, we checked the size of this effect by generating models with an exponentially declining star-formation
history (e-folding timescale $\tau=1$~Gyr) with various stellar metallicities, using both Salpeter and BG03 IMFs (without extinction). We
determined that ages using the (top-heavy) BG03 IMF are $\sim40-50\%$ larger 
for galaxies which are found to be $\sim1$~Gyr old using a Salpeter IMF. (Note that 
derived metallicities using the BG03 IMF are larger too).

\begin{figure*}[htp]
\epsscale{1.5}\plotone{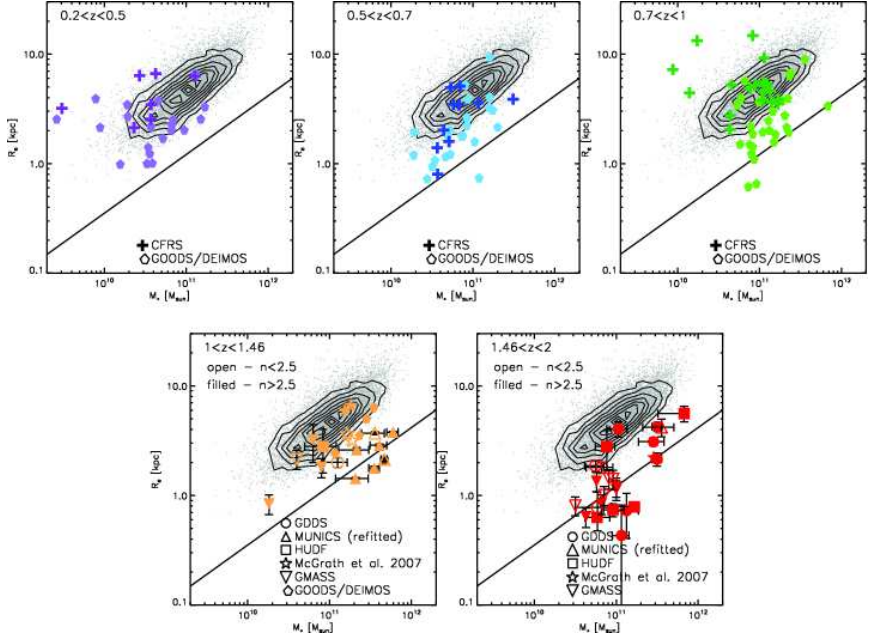}
\caption{As for Figure~\ref{f5}, with data from the GOODS\slash DEIMOS and CFRS
redshift surveys included. Points corresponding to different redshift bins are presented in separate panels. The solid line is the best-fit relation from 
Figure~\ref{f5}.}
\label{f8}
\end{figure*}

\begin{figure*}[htp]
\epsscale{1.5}\plotone{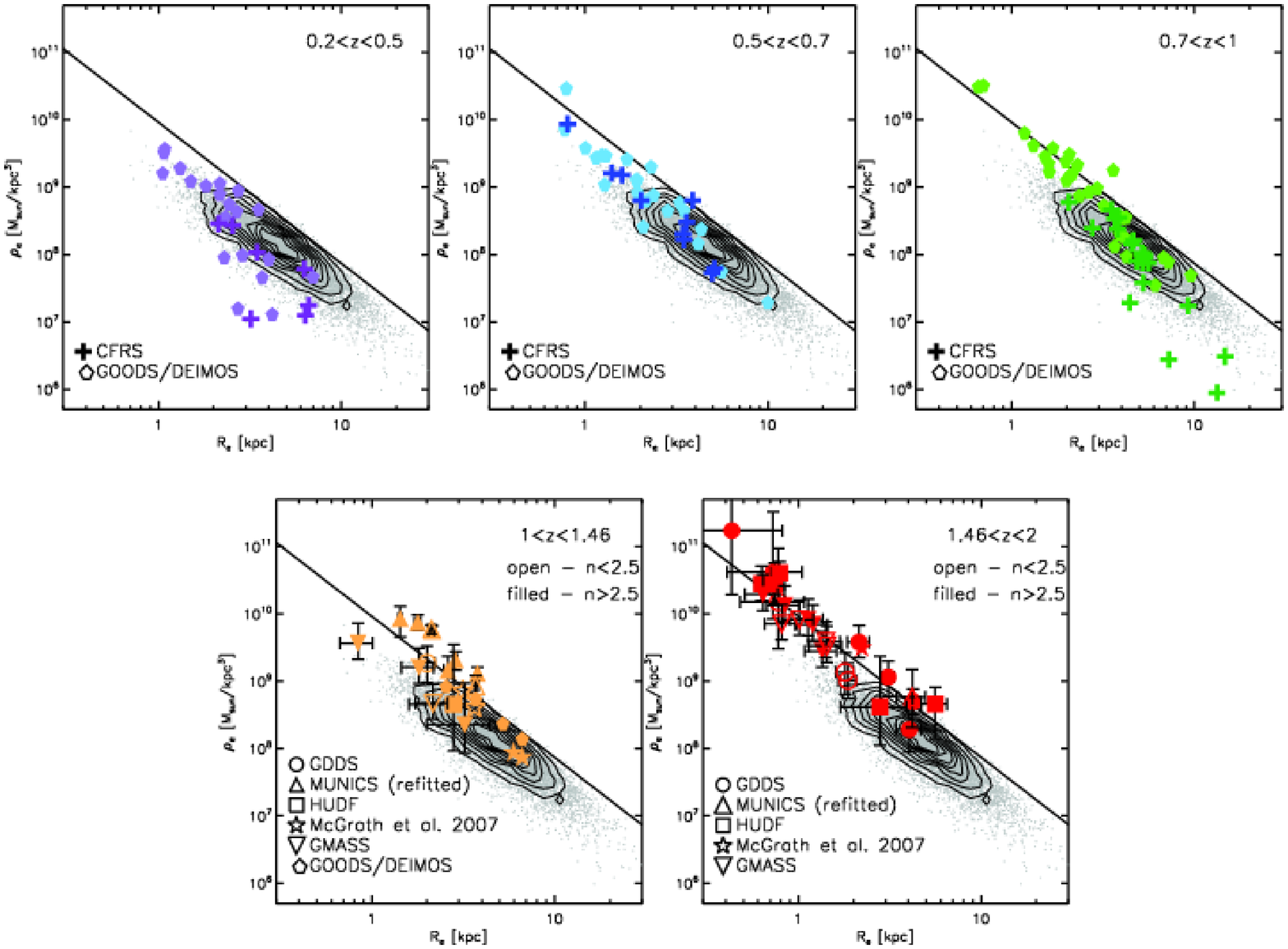}
\caption{As for Figure~\ref{f6}, with data from the GOODS\slash DEIMOS and CFRS
redshift surveys included. Points corresponding to different redshift bins are presented in separate panels. The solid line is the best-fit relation from 
Figure~\ref{f6}.}
\label{f9}
\end{figure*}

\begin{figure*}[htp]
\epsscale{2}\plotone{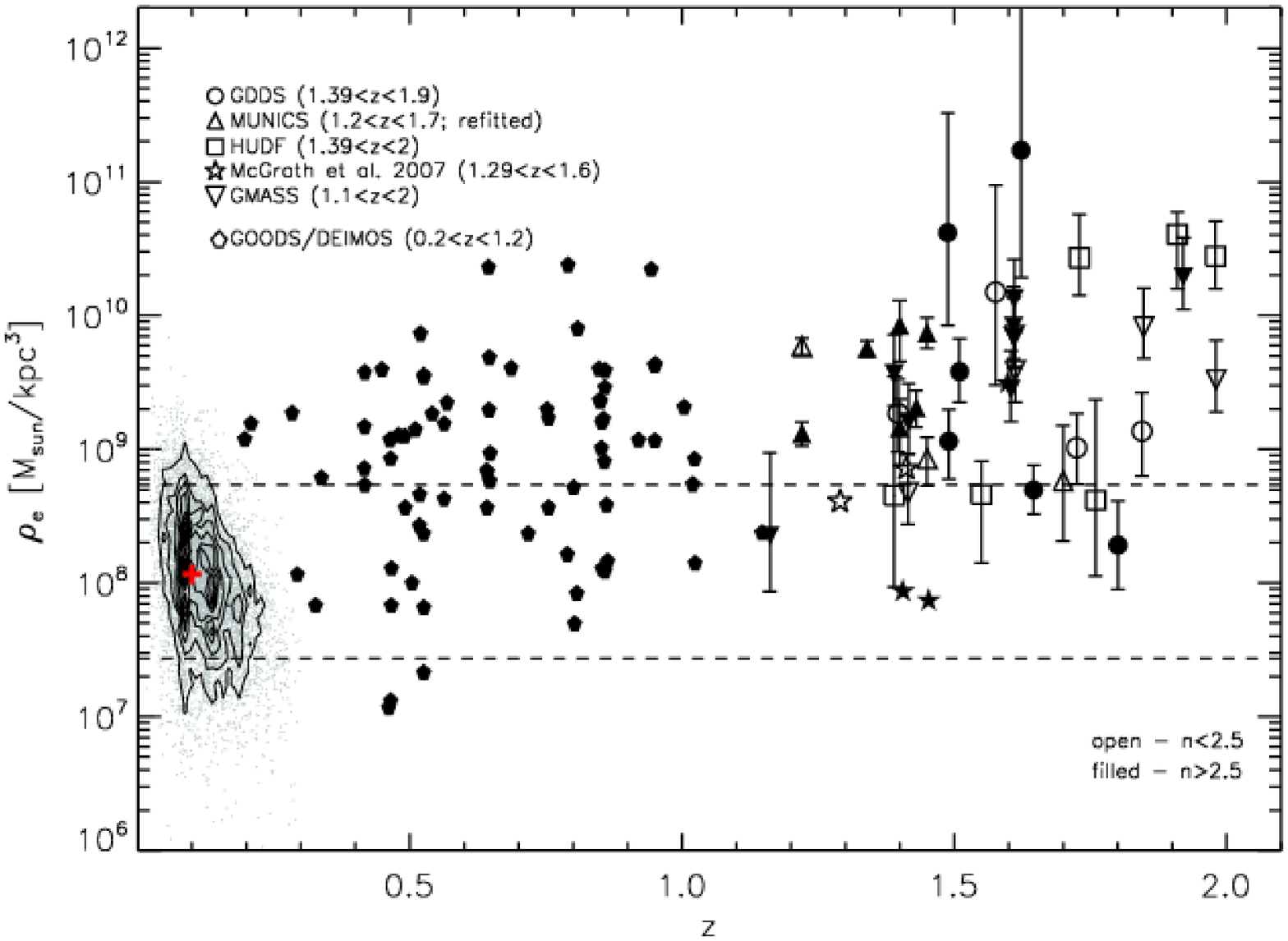}
\caption{Redshift dependence of the stellar mass density within the effective radius $R_e$. Symbols are as in Figures~\ref{f8} and~\ref{f9}. The local sample of SDSS galaxies is presented with both points and overlaid contours that denote linearly spaced regions of constant density of galaxies in this parameter space. Red cross represents the median stellar mass density and the median redshift of the local sample. Limiting stellar mass densities for the 90th percentiles for the SDSS objects with stellar mass densities above and below the median value are given with upper and lower dashed lines, respectively. Following the  discussion on the quality of CFRS imaging in section~\ref{disc} corresponding points are excluded from this fugure.}
\label{f10}
\end{figure*}

Some constraints on
the duty cycle for the size change can be inferred from our observations,
by noting that the redshift range spanned
by our sample is $1.1<z<2.0$, corresponding to 
a spread in time of $\sim2.2$~Gyr. The 
division of the sample
in half at $z=1.46$ using different
symbols in Figures~\ref{f5} and~\ref{f6}
subdivides this redshift
interval into two equal time bins, each of which
is $\sim1.1$~Gyr wide.
The  sample shown in Figures~\ref{f5} and~\ref{f6} contains data
from a number of different surveys, and it is certainly
unwise to attempt to compare the
high-redshift and low-redshift subsets 
at a detailed level. But it is perhaps
worth noting the following
very general qualitative
trends.
Figure~\ref{f5} appears to show that the
character of the size-mass distribution
is rather
different in the $1.1<z<1.46$ and
$1.46<z<2.0$
intervals, 
with neither distribution resembling the local 
data distribution closely. This suggests
some degree of evolution between the 
bins, but with the caveat that these two redshift 
bins primarily consist of data from different surveys so the strength 
of  the evolution cannot be confidently inferred. 
On a more speculative note, it can be argued
that nothing in Figure~\ref{f5} rules
out the possibility that the high-redshift distribution
is evolving into the low-redshift distribution differentially,
with different physics operating at the low mass and
high mass ends.
In fact, some evidence for this is also
hinted at 
in the figure, which appears to
show that the
smallest and least massive galaxies lie
at $z>1.5$.
It is possible that dry mergers may well
be growing the smallest and least massive galaxies
along the fundamental plane early in a galaxy's
life cycle, before some other process takes over
and grows them further in some other way.

It is interesting to contrast the data presented
in Figures~\ref{f5} and~\ref{f6} with data which spans 
the redshift range in between the SDSS 
data and our high-redshift observations. 
Figures~\ref{f8} and~\ref{f9} 
augment the data in Figures~\ref{f5} and~\ref{f6} with
intermediate-redshift
data taken from the CFRS \citep{sch99, lil95}
and GOODS\slash DEIMOS \citep{bte07,treu05} surveys. 
Effective radii for the CFRS objects are obtained from the WFPC2 814W images. 
Estimates based on the images in three ACS filters 
(606W, 814W, and 850W) are available for the 
GOODS/DEIMOS sample. All objects shown in the upper three panels  
of figures~\ref{f8} and~\ref{f9} have sizes based on 
the WFPC2 or ACS 814W imaging that translates 
approximately into the rest-frame $V$-band for the 
median redshifts in the $0.2<z<0.5$ and $0.5<z<0.7$ panels, 
and into the rest-frame $B$-band for the median redshift in 
the $0.7<z<1.$ panel. GOODS/DEIMOS objects in 
the $1<z<1.46$ panel are presented with the effective radii 
in ACS 850W filter (approximately $B$-band rest frame). 
The CFRS masses are obtained following \citet{bal08} and using imaging
data of relatively low quality. We note that the difference in the 
rest-frame wavelengths that are probed at different redshifts makes 
it impssible to draw any quantitative conclusions about galaxy size evolution. 
However, figures~\ref{f8} and~\ref{f9} show qualitative trends consistent 
with smooth evolution over the $0.2<z<2$ range. The dispersion
on the size-mass plot in the $0.2<z<1$ regime is large (upper panels in Figures~\ref{f8} and~\ref{f9}), but there seems 
to be some evidence for a systematic offest relative to the local
trends with the increasing redshift. The GOODS\slash DEIMOS data in Figures~\ref{f8} and~\ref{f9} spans {\em both}
the low-redshift and high-redshift loci identified
in  each panel of Figures~\ref{f8} and~\ref{f9} by contours and the line of the best fit, respectively. However, the majority of the
low-redshift ($0.2<z<0.5$) GOODS\slash DEIMOS data lie closer
to the local relation, in contrast to the $0.7<z<1$ panel where the most of the GOODS\slash DEIMOS points are close to the $z\sim1.5$ objects locus. 
The CFRS data in the $0.7<z<1$ panel of Figure~\ref{f8} does not seem to follow this trend, and we suggest that it may be due to the
shallow imaging of these objects \citep{lil95}. In the lower redshift panels of  Figure~\ref{f8} ($z<0.7$) the positions of the CFRS objects are 
consistent with the GOODS\slash DEIMOS dataset. In order to compare the number of high 
mass objects at different redshifts, we use a subsample of 68 GOODS/DEIMOS objects with 
masses above the GDDS detection limit (see section~\ref{sdef}). 
It is interesting to note that relatively few ($14\slash 68$, $\sim21\%$) points
from the GOODS\slash DEIMOS subsample have masses
greater than $1.5\times10^{11}$~M$_{\sun}$ \citep[M$^{\ast}$ at $1<z<2$,][]{fon06}.
In contrast, the high-redshift data set presented in Figures~\ref{f5} and~\ref{f6} includes large fraction
of objects with M$_{\star}>$~M$^{\ast}$ -- $18\slash 43$ ($\sim42\%$). 
While this could perhaps be consistent with
adiabatic mass loss,  the arguments
presented in our discussion of Figure~\ref{f7} 
are compounded by the data presented in Figures~\ref{f8} and~\ref{f9}
which indicates that size growth is still occurring
in galaxies even older than those in our GDDS
sample. We think it is likely that the absence of very high
mass objects in the GOODS\slash DEIMOS data is simply 
due subtle differences in various groups'
methodologies for computing stellar masses
from photometric data. To further address the question of structural evolution of galaxies presented in Figures~\ref{f8} and~\ref{f9} we plot the redshift 
dependence of the projected stellar mass density (defined in section~\ref{sizemass}) in Figure~\ref{f10}. Dashed lines encompass the range of mass 
density which contains 90\% of the local (SDSS) data points. The median stellar mass density of the SDSS galaxies is $\rho_e=1.1\times10^8$M$_{\sun}$~kpc$^{-3}$ and this value is 
denoted by a red cross plotted at $z=0.1$ in Figure~\ref{f10}. Large fraction (88\%) of the GOODS\slash DEIMOS objects have mass densities above the local median value, and 65\% of these galaxies have mass densities above the upper dashed line in the figure. For the $1.1<z<2$ sample the corresponding numbers are 90\% and 77\%, respectively. On this basis we can conclude that the stellar mass density increases over an extended redshift range, though the dispersion of the plot is large, and more points in both intermediate and high redshift regime are needed to properly constrain this redshift dependence. We intend to revisit the topic in a future paper.

On balance, we conclude that at present neither
adiabatic expansion nor equal-mass dry mergers
seem able to explain the size growth in early-type galaxies. 
A successful model will have to simultaneously explain
the size change in the galaxies, the duty cycle
for this size change, and the epoch in a galaxy's
life history at which the change occurs. 
And, as noted above, mass density growth over
the redshift interval being probed suggests
that the size growth 
being witnessed is operating within a broader
context for galaxy formation.
Over the redshift interval where early-type
galaxies are growing in size, the
volume-averaged stellar
mass density in massive
galaxies is increasing, and
the morphological mix is changing.

\section{Conclusions}
The size-mass
relationship for early-type
galaxies evolves significantly from $z=2$ to $z=1$.
Over the whole of this
redshift range early type galaxies tend to be a factor
of $2-3$ smaller
than local counterparts of similar mass. Similarly
compact galaxies are seen at $z>2$  \citep{van08}, and
we speculate that
the very compact galaxies studied in the present
paper are simply the evolved counterparts of these higher-redshift objects,
caught at a time before subsequent size growth. By comparing
the size distribution of our sample with that of lower
redshift surveys, we conclude that significant
size growth is probably occurring over the redshift range explored
in the present paper.  The physics of this growth remains
mysterious. By comparing the size-mass relation at $z\sim1.5$ with its local
counterpart we conclude that equal mass dry mergers play only a limited role
in growing early-type galaxies, at least once they are
older than a few~Gyr.  Other processes
may be as important as dry merging in growing early-type galaxies.
Adiabatic expansion is one such process that we have
examined, and while it may be
important in growing young early-type galaxies,
it is hard to see how this mechanism can be invoked
to obtain a factor of two
growth in the sizes of galaxies
as old as those in the present survey.

\acknowledgements
\noindent{\bf Acknowledgments}
\bigskip

\noindent We thank Norm Murray for generously sharing his ideas
and papers in advance of publication. We also thank Kevin Bundy
for useful discussions.

This paper is based on observations obtained at the Gemini
Observatory, which is operated by the Association of Universities for
Research in Astronomy, Inc., under a cooperative agreement with the
NSF on behalf of the Gemini partnership: the National Science
Foundation (United States), the Particle Physics and Astronomy
Research Council (United Kingdom), the National Research Council
(Canada), CONICYT (Chile), the Australian Research Council
(Australia), CNPq (Brazil) and CONICET (Argentina).

Based on observations made with the NASA\slash ESA Hubble Space Telescope, obtained at the Space Telescope Science Institute, which is operated by the Association of Universities for Research in Astronomy, Inc., under NASA contract NAS 5-26555. These observations are associated with program \#9760. Support for program \#9760 was provided by NASA through a grant from the Space Telescope Science Institute, which is operated by the Association of Universities for Research in Astronomy, Inc., under NASA contract NAS 5-26555.

RGA thanks 
NSERC, the Government of Ontario, and the Canada Foundation for
Innovation
for funding provided by an E. W. R. Steacie Memorial Fellowship.


\begin{thebibliography}

\bibitem[Abraham et al.(2004)]{abr04} Abraham, R.~G., et al.\ 2004, \aj, 127, 2455 (Paper~I)
\bibitem[Abraham et al.(2007)]{abr07} Abraham, R.~G., et al.\ 2007, \apj, 669, 184 (Paper~VIII)
\bibitem[Baldry \& Glazebrook(2003)]{bal03} Baldry, I.~K., \& Glazebrook, K.\ 2003, \apj, 593, 258
\bibitem[Baldry et al.(2008)]{bal08} Baldry, I.~K., Glazebrook, K., \& Driver, S.~P.\ 2008, \mnras (in press), arXiv:0804.2892
\bibitem[Baumgardt et al.(2007)] {bk07}Baumgardt, H., \& Kroupa, P. 2007, MNRAS, 380, 1589 
\bibitem[Bell et al.(2004)]{bell04} Bell, E.~F., et al.\ 2004, \apj, 608, 752
\bibitem[Bell et al.(2006a)]{bell06a} Bell, E.~F., et al.\ 2006a, \apj, 652, 270
\bibitem[Bell et al.(2006b)]{bell06b} Bell, E.~F., et al.\ 2006b, \apj, 640, 241 
\bibitem[Bernardi et al.(2003)]{ber03} Bernardi, M., et al.\ 2003, \aj, 125, 1817
\bibitem[Binney  \& Merrifield(1998)]{1998gaas.book.....B} Binney, J., \& Merrifield, M.\ 1998, Galactic astronomy \slash ~James Binney and Michael Merrifield.~ Princeton, NJ : Princeton University Press, 1998.~ (Princeton series in astrophysics) QB857 .B522 1998
\bibitem[Boylan-Kolchin et al.(2005)]{bk05} Boylan-Kolchin, M., Ma, C., \& Quataert, E.\ 2005, \mnras, 362,184
\bibitem[Boylan-Kolchin et al.(2006)]{bk06} Boylan-Kolchin, M., Ma, C., \& Quataert, E.\ 2006, \mnras, 369,1081
\bibitem[Brown et al.(2007)]{brown07} Brown, M. J. L., Dey, A., Jannuzi, B. T., Brand, K., Benson, A., Brodwin, M. Croton D. J., Eisenhardt, P. R. 2007, \apj, 654, 858
\bibitem[Brown et al.(2003)]{brown03} Brown, M. J. L., Dey, A., Jannuzi, B. T., Lauer, T. R., Tiede, G., Mikles, V. J. 2003, \apj, 597, 225
\bibitem[Bruzual \& Charlot(2003)]{bc03} Bruzual, G., \& Charlot, S.\ 2003, \mnras, 334, 1000
\bibitem[Bundy et al.(2007)]{bte07} Bundy, K., Treu, T., \& Ellis, R.~S.\ 2007, \apj, 665, L5
\bibitem[Charlot et al.(1996)]{cha96} Charlot, S., Worthey, G., \& Bressan, A. 1996, \apj, 457, 625
\bibitem[Chen et al.(2003)]{chen03} Chen, H.-W., Marzke, R., McCarthy, P., Martini, P., Carlberg, R. et al. 2003, \apj, 586, 745
\bibitem[Cimatti et al.(2002)]{cim02} Cimatti, A., et al.\ 2002, \aap, 381, L68 
\bibitem[Cimatti et al.(2004)]{cim04} Cimatti, A., et al.\ 2004, \nat, 430, 184 
\bibitem[Cimatti et al.(2008)]{cim08} Cimatti, A., et al.\ 2008, \aap, 482, 21
\bibitem[Daddi et al.(2004)]{dad04} Daddi, E., et al.\ 2004, \apj, 600, 127
\bibitem[Daddi et al.(2005a)]{dad05a} Daddi, E., et al.\ 2005, \apj, 631, L13
\bibitem[Daddi et al.(2005b)]{dad05b} Daddi, E., et al.\ 2005, \apj, 626, 680
\bibitem[De Lucia \& Blaizot(2007)]{dlb07} De Lucia, G., \& Blaizot, J.\ 2007, \mnras, 375, 2
\bibitem[Dickinson et al.(2003)]{dic03} Dickinson, M. et al. 2003, \apj, 587, 25
\bibitem[Drory et al.(2001)]{dr01} Drory, N., et al.\ 2001, \mnras, 325, 550
\bibitem[Eggen, Lynden-Bell \& Sandage(1962)]{elbs62} Eggen, O.~J., Lynden-Bell, D., \& Sandage, A.~R.\ 1962, \apj, 136, 748
\bibitem[Faber \& Jackson(1976)]{FJ76} Faber S. M., Jackson R. E., 1976, ApJ, 204,  668 
\bibitem[Faber et al.(2007)]{fab07} Faber, S.~M., et al.\ 2007, \apj, 665, 265
\bibitem[Fontana et al.(2004)]{fon04} Fontana, A., et al.\ 2004, \aap, 424, 23
\bibitem[Fontana et al.(2006)]{fon06} Fontana, A., et al. 2006, \aap, 459, 745
\bibitem[Fruchter \& Hook(2002)]{fh02} Fruchter, A.~S., \& Hook, R.~N.\ 2002, \pasp, 114, 144
\bibitem[Garmany \& Conti(1985)]{gar85} Garmany, C.D. \& Conti, P.S. 1985, \apj, 293, 407
\bibitem[Gallagher, Hunter \& Tutukov(1984)]{ght84} Gallagher, J.~S., Hunter, D.~A., \& Tutukov, A.~V.\ 1984, \apj, 284, 544
\bibitem[Glazebrook et al.(2004)]{gla04} Glazebrook, K., et al.\ 2004, \nat, 430, 181 (Paper~III)
\bibitem[Gonz\'alez-Garc\'ia \& van Albada(2003)]{gva03} Gonz\'alez-Garc\'ia, A.~C., \& van Albada, T.~S.\ 2003, \mnras, 342, 36
\bibitem[Hills(1980)]{hi80}Hills, J.~G.\ 1980, \apj, 225, 986.
\bibitem[Jedrzejewski(1987)]{jed87} Jedrzejewski, R. I., 1987, \mnras, 226, 747
\bibitem[J\o rgensen et al.(1995)]{jor95} J\o rgensen, I., Franx, M., \& Kj\ae rgaard, P.\ 1995, \mnras, 273, 1097
\bibitem[Kang, Jing, \& Silk(2006)]{KJS06} Kang X., Jing Y. P., Silk J., 2006, ApJ, 648,  820 
\bibitem[Kauffmann et al.(2003)]{kau03} Kauffmann, G., et al.\ 2003, \mnras, 341, 33
\bibitem[Kormendy(1977)]{K77} Kormendy J., 1977, ApJ, 218,  333 
\bibitem[Kriek et al.(2006)]{kr06} Kriek, M., et al.\ 2006, \apj, 649, L71 
\bibitem[Kriek et al.(2007)]{kr07} Kriek, M. et al.\ 2007, \apj, 669, 776 
\bibitem[Labb{\'e} et al.(2005)]{la05} Labb{\'e}, I., et al.\ 2005, \apj, 624, L81 
\bibitem[Le Borgne et al.(2004)]{lb04}Le Borgne, D., Rocca-Volmerange, B.,Prugniel, P., Lanon, A., Fioc, M., Soubiran, C.\ 2004, \aap,  425, 881
\bibitem[Lilly et al.(1995)]{lil95} Lilly, S.~J., Le F\'evre, O., Crampton, D., Hammer, F., \& Tresse, L.\ 1995, \apj, 455, 50
\bibitem[Longhetti et al.(2005)]{lon05} Longhetti, M., et al.\ 2005, \mnras, 361, 897
\bibitem[Longhetti et al.(2007)]{lon07} Longhetti, M., et al.\ 2007, \mnras, 374, 614
\bibitem[Maraston et al.(2006)]{mar06} Maraston, C., et al.\ 2006, \apj, 652, 85 
\bibitem[McCarthy et al.(2001)]{mcc01} McCarthy, P.~J., Carlberg, R., Chen, H.-W., Marzke, R., Firth, A., et al. 2001, \apj, 560, L11
\bibitem[McCarthy et al.(2004)]{mcc04} McCarthy, P.~J., et al.\ 2004, \apj, 614, L9 (Paper~IV)
\bibitem[McCarthy et al.(2007)]{mcc07} McCarthy, P.~J., et al.\ 2007, \apj, 664, L17 
\bibitem[McGrath et al.(2007a)]{mcg07a} McGrath, E.~J., Stockton, A., \& Canalizo, G.\ 2007, \apj, 669, 241
\bibitem[McGrath et al.(2007b)]{mcg07b} McGrath, E.~ J., et al.\ 2007, \apj (submitted),  arXiv: 0707.1050
\bibitem[Moustakas et al.(2004)]{mou04} Moustakas, L.~A., et al.\ 2004, \apj, 600, L131
\bibitem[Murray, Quataert \& Thompson(2008, in preparation)]{mu08} Murray, N., Quataert, E., \& Thompson, T.~A.\  2008 (in preparation)
\bibitem[Naab et al.(2007)]{na07} Naab, T., Johansson, P.~H., Ostriker, J.~P., \& Efstathiou, G.\ 2007, \apj, 658, 710
\bibitem[Peng et al.(2002)]{pe02} Peng, C.~Y.,  Ho, L.~C., Impey, C.~D., \& Rix, H.-W.,\ 2002, \aj, 124, 266
\bibitem[Pipino and Matteucci(2008)]{pip08}Pipino, A. \& Matteucci, F. 2008, A\&A (in press), arXiv:0805.0793
\bibitem[Rudnick et al.(2003)]{ru03} Rudnick, G, et al.\ 2003, \apj, 599,847
\bibitem[Rudnick et al.(2006)]{ru06} Rudnick, G. et al.\ 2006, \apj, 650, 624
\bibitem[Schade et al.(1999)]{sch99} Schade, D., et al.\ 1999, \apj, 525, 31
\bibitem[Schombert (1986)]{schombert86} Schombert, J. M. 1986, \apjs, 60, 602
\bibitem[Schweizer (1987)]{schweizer87} Schweizer, F. 1987, Science, 231, 227 
\bibitem[Searle \& Zinn(1978)]{searle78} Searle, L., Zinn, R. 1978, \apj, 223, 82 
\bibitem[Stockton et al.(2004)]{sto04} Stockton, A., Canalizo, G., \& Maihara, T.\ 2004, \apj, 605, 37 
\bibitem[Tody (1993)]{tody93} Tody, D., 1993, ASP Conf. Series, 52, 173. 
\bibitem[Toomre \& Toomre(1972)]{tt72} Toomre, A., \& Toomre, J.\ 1972, \apj, 178, 623
\bibitem[Toft et al.(2007)]{to07} Toft, S., et al.\ 2007, \apj, 671, 285
\bibitem[Treu et al.(2005)]{treu05} Treu, T., et al. 2005, \apj, 633, 174
\bibitem[van Dokkum et al.(2008)]{van08} van Dokkum, P.~G., et al.\ 2008, \apj, 677, L5
\bibitem[van Dokkum(2005)]{van05} van Dokkum, P.~G.\ 2005, \aj, 130, 2647
\bibitem[Yan \& Thompson(2003)]{yan03} Yan, L., \& Thompson, D.\ 2003, \apj, 586, 765
\bibitem[Zhao(2002)]{zh02} Zhao, H.~S.\ 2002, \mnras, 336, 159 
\bibitem[Zirm et al.(2003)]{zi03} Zirm, A.~W., Dickinson, M., \& Dey, A.\ 2003, \apj, 585, 90
\end{thebibliography}
\end{document}